\documentclass[10pt,twocolumn,twoside,journal,final]{IEEEtran}
\usepackage{amssymb}
\usepackage{xcolor}
\usepackage{soul}
\usepackage{siunitx}
\usepackage{overpic}
\usepackage{rotating}
\usepackage[export]{adjustbox}
\usepackage{pifont}

\usepackage{tikz}
\usetikzlibrary{calc}
\usepackage{todonotes}
\usepackage{enumitem}

\ifCLASSOPTIONdraftcls
\usepackage[colorlinks=true,linkcolor=violet]{hyperref}
\fi

\usepackage[numbers,sort]{natbib}

\usepackage{cite}
\usepackage{mathtools}

\usepackage{accents} %for scalable overset arrows e.g.  $\accentset{\rightharpoonup}{#1}{abcd}$
\usepackage{harpoon}%

\newcommand{\sectionoutline}[2][]{}

\ifCLASSOPTIONdraftcls
\usepackage[left=2cm,right=7cm,top=3cm,bottom=3cm]{geometry}
\usepackage{showlabels}
\fi

\usepackage{printlen} %can be used to print out useful lengths
\usepackage{relsize}
\usepackage[printonlyused,withpage]{acronym} % add 'smaller' option to make small caps

\ifCLASSINFOpdf
\else
\fi
\usepackage{amsmath}
\usepackage[noend]{algpseudocode} %loads algorihmicx as well
\ifCLASSOPTIONcompsoc
 \usepackage[caption=false,font=normalsize,labelfont=sf,textfont=sf]{subfig}
\else
 \usepackage[caption=false,font=footnotesize]{subfig}
\fi
\usepackage{float}
\usepackage{siunitx}

\usepackage{url}
\usepackage[colorlinks]{hyperref}
\usepackage[symbols,nogroupskip,sort=none]{glossaries-extra}

\newcommand{\Iso}{_\text{Iso}}
\newcommand{\SSIso}{_\text{SS-Iso}}

\newcommand{\Hyb}{_\text{Hyb}}
\newcommand{\SSHyb}{_\text{SS-Hyb}}
\newcommand{\Out}{_\text{z}}
\newcommand{\SubS}{_\text{SS}}
\newcommand{\Oracle}{_\text{gt}}
\newcommand{\SSH}{\text{SSH}}
\newcommand{\SSHK}{\text{SSH-K}}

\newcommand{\Aniso}{_\text{Aniso}}
\newcommand{\Identity}{_\text{Identity}}
\newcommand{\uPeak}{_\text{p}} % subscript for unimodality peak
\newcommand{\PW}{_\text{PW}}
\newcommand{\Target}{_\text{s}}

\newcommand{\iMic}{q}
\newcommand{\nMic}{Q}
\newcommand{\direction}{\Omega}
\newcommand{\az}{\varphi}
\newcommand{\inc}{\theta}
\newcommand{\iFreq}{\nu}
\newcommand{\iFrame}{\ell}
\newcommand{\nFreq}{F}

\newcommand{\TF}{\iFreq,\iFrame}

\newcommand{\iPW}{j}
\newcommand{\nPW}{J}
\newcommand{\dict}{\mathcal{D}}
\newcommand{\iModel}{m}
\newcommand{\nModel}{M}
\newcommand{\nDir}{\Psi}
\newcommand{\nCluster}{K}
\newcommand{\nData}{N}
\newcommand{\nTalker}{N_{s}}
\newcommand{\timeConst}{T}

\newcommand{\stftWeightVec}{\mathbf{w}}
\newcommand{\stftArray}{Y}
\newcommand{\stftArrayVec}{\mathbf{y}}
\newcommand{\ATF}{\mathbf{h}}
\newcommand{\steerVec}{\mathbf{d}}
\newcommand{\stftPW}{X}
\newcommand{\stftPWVec}{\mathbf{x}}
\newcommand{\stftNoiseVec}{\mathbf{n}}
\newcommand{\stftTarget}{S}
\newcommand{\stftTargetVec}{\mathbf{s}}
\newcommand{\stftOut}{Z}
\newcommand{\stftOutVec}{\mathbf{z}}
\newcommand{\stftOutMat}{\mathbf{Z}}
\newcommand{\stftWhite}{V}
\newcommand{\stftWhiteVec}{\mathbf{v}}
\newcommand{\quadWeight}{q}
\newcommand{\covRSet}{\mathfrak{R}}
\newcommand{\covrSet}{\mathfrak{r}}
\newcommand{\DOAset}{\mathcal{I}}
\newcommand{\SteeringSet}{\DOAset_{s}}
\newcommand{\smoothness}{\alpha}
\newcommand{\EigenVec}{\mathbf{U}}
\newcommand{\EigenVal}{\mathbf{\Sigma}}
\newcommand{\EigenVecVal}{U}
\newcommand{\AllOnesMat}{\mathbf{J}}
\newcommand{\AllOnesVec}{\mathbf{j}}
\newcommand{\signatureMat}{\mathbf{K}}

\newcommand{\sigmaPlus}{\mathbf{\sigma}_{1}}
\newcommand{\sigmaMinus}{\mathbf{\sigma}_{2}}

\newcommand{\covMat}{\mathbf{R}}
\newcommand{\covVec}{\mathbf{r}}
\newcommand{\scalarLoading}{\varepsilon}
\newcommand{\covRobust}{\covMat_{\scalarLoading}}
\newcommand{\covNCM}{\hat{\covMat}_{\mathcal{N}}}
\newcommand{\covEigenvalue}{\lambda}
\newcommand{\sigMicDiffuseFreqVec}{\mathbf{\gamma}}
\newcommand{\covDiffuse}{\covMat_{\sigMicDiffuseFreqVec}}
\newcommand{\covEstIso}{\covMat\Iso}
\newcommand{\covEstOracle}{\covMat\Oracle}

\newcommand{\isotropy}{\mathcal{P}}
\newcommand{\powDyn}{\mathcal{A}}

\newcommand{\hermitianOp}{^{H}}
\newcommand{\transposeOp}{^{T}}
\newcommand{\expectOp}{\mathbb{E}}
\newcommand{\eye}{\ensuremath \mathbf{I}}
\DeclareMathOperator{\diag}{diag}

\DeclareMathOperator*{\argmin}{arg\,min}

\renewcommand{\deg}{\ensuremath ^{\circ}}

\newcommand{\timeconst}{\alpha}
\newcommand{\Complex}{\mathbb{C}}
\newcommand{\Real}{\mathbb{R}}
\newcommand{\covCondNumLimit}{\kappa_{0}}

\usepackage[]{preview}
\usepackage{nameref}

\ifPreview

\makeatletter\renewcommand\@eqnnum{}\makeatother

\makeatletter
\let\@eqnswtrue\@eqnswfalse
\makeatother

\definecolor{ic_text_grey}{RGB}{75,79,85}

\PreviewEnvironment{align}

\usepackage{etoolbox}
\AtBeginEnvironment{math}{\color{ic_text_grey}}
\AtBeginEnvironment{equation}{\color{ic_text_grey}}
\AtBeginEnvironment{equation*}{\color{ic_text_grey}}
\AtBeginEnvironment{eqnarray}{\color{ic_text_grey}}
\AtBeginEnvironment{eqnarray*}{\color{ic_text_grey}}
\AtBeginEnvironment{alignat}{\color{ic_text_grey}}
\AtBeginEnvironment{alignat*}{\color{ic_text_grey}}
\AtBeginEnvironment{multline}{\color{ic_text_grey}}
\AtBeginEnvironment{multline*}{\color{ic_text_grey}}
\AtBeginEnvironment{align}{\color{ic_text_grey}}
\AtBeginEnvironment{align*}{\color{ic_text_grey}}
\everymath{\color{ic_text_grey}}
\fi
\begin{document}

\newcommand{\acro}{\acrodef}\newcommand{\acroindefinite}{\acrodefindefinite}%% List of SAP acronyms. Some non-obvious points to note are:
%% (1) Keep in alphabetical order to print correctly.
%%     The perl script "tidy_acronyms.pl" will sort the acronyms.
%% (2) Put each complete acronym definition in a single line
%% (3) No non-standard characters. Accents should use the LaTeX escapes.
%%     The perl script "tidy_acronyms.pl" will replace accents with the correct form.
%% (4) Use \acs{...} to include another acronym in the definition
%%     e.g. \acro{SSNR}{Segmental \acs{SNR}}
%% (5) Text additional to the definition should be enclosed thus: \acroextra{ }
%%     e.g. \acro{AAC}{Advanced Audio Coding\acroextra{. A lossy codec used for digital audio.}}
%% (6) Multiple meanings of an acronym are distinguished thus:
%%     e.g. \acro{SNR1}[SNR]{Society for Nautical Research}
%% (7) You must also use this form if the the acronym involves mathematical symbols:
%%     e.g. \acro{Leq}[L$_{\textrm{eq}}$]{Equivalent Continuous Sound Level}
%% (8) Non-standard plurals can be defined as:
%%     e.g. \acro{MP}{Members of Parliament}\acrodefplural{MP}[MPs]{Members of Parliament}
%% (9) Different indefinite articles can be specified for long and short forms
%%     e.g. \acro{FFT}{Fast Fourier Transform}\acroindefinite{FFT}{an}{a}
%%
\acro{AR}{augmented reality} 
\acro{AEC}{acoustic echo control} 
\acro{AI}{artificial intelligence}
\acro{AIR}{acoustic impulse response}
\acro{ATF}{acoustic transfer function}
\acro{ASR}{automatic speech recognition} 
\acro{A-SLAM}{acoustic simultaneous localization and mapping}
\acro{BSE}{blind signal extraction}
\acro{BTE}{behind the ear}
\acro{BSS}{blind source separation}
\acro{CDR}{coherent to diffuse ratio}
\acro{CM}{compact model}
\acro{DAS}{delay-and-sum}
\acro{DBSTOI}{binaural \acs{STOI}}
\acro{DL}{deep learning}
\acro{DNN}{deep neural network}
\acro{DOA}{direction of arrival}
\acrodefplural{DOA}[DOAs]{directions of arrival}
\acro{DOF}{degree of freedom}
\acrodefplural{DOF}{degrees of freedom}
\acro{DTFT}{discrete time Fourier transform}
\acro{EMA}{exponential moving average}
\acro{EVD}{Eigenvalue Decomposition}
\acro{EWLS}{exponentially-weighted least squares}
\acro{FIR}{finite impulse response\acroextra{. A filter whose output is a weighted sum of past input values and whose system function contains only zeros and no poles.}}
\acro{FSB}{filter-and-sum beamformer}
\acro{fwSegSNR}{frequency-weighted segmental \ac{SNR}}
\acro{GSC}{generalized sidelobe canceller}
\acro{GSS}{geometric source separation} 
\acro{GHRTF}{generalized head-related transfer function}
\acro{GWPE}{generalized weighted prediction error}
\acro{HA}{hearing aid}
\acro{HAHRIR}{hearing aid head-related impulse response}
\acro{HAHRTF}{hearing aid head-related transfer function}
\acro{HARIR}{hearing aid room impulse response}
\acro{HATS}{head and torso simulator}
\acro{HRIR}{head-related impulse response}
\acro{HRTF}{head-related transfer function}
\acro{HSWOBM}{high-resolution stochastic \acs{WSTOI}-optimal binary mask}
\acro{ICA}{independent component analysis}
\acro{ITD}{interaural time difference}
\acro{ILD}{interaural level difference}
\acro{IMU}{inertial measurement unit}
\acro{IR}{impulse response}
\acro{ISFT}{inverse \ac{SFT}}
\acro{KKT}{Karush-Kuhn-Tucker}
\acro{LCMV}{linearly constrained minimum variance}
\acro{LTASS}{long term average speech spectrum}
\acro{LTI}{linear time-invariant}
\acro{ML}{maximum likelihood}
\acro{MISO}{multiple-input single-output}
\acro{SISO}{single-input single-output}
\acro{MMSE}{minimum mean squared error}
\acro{MWF}{multichannel Wiener filter}
\acro{MPDR}{minimum power distortionless response}\acroextra{ beamformer}
\acro{MVDR}{minimum variance distortionless response}\acroextra{ beamformer}
\acro{MVDR}{minimum variance distortionless response\acroextra{ beamformer}}
\acro{NCM}{noise covariance matrix}
\acro{NMF}{non-negative matrix factorization}
\acro{NPD}{noise power distribution}
\acro{PIV}{pseudointensity vector}
\acro{PESQ}{perceptual evaluation of speech quality}
\acro{PCA}{principal component analysis}
\acro{PSD}{power spectral density}
\acro{PW}{plane-wave}
\acro{PWD}{plane-wave density}
\acro{PWDPSD}{\ac{PWD}\ac{PSD}}
\acro{RIR}{room impulse response}
\acro{RLS}{recursive least squares}
\acro{RLSFI}{robust least-squares frequency-invariant}
\acro{RMS}{root mean square}
\acro{RS}{recursive smoothing}
\acro{RT}{reverberation time}
\acro{RTF}{relative transfer function}
\acro{SAD}{speech activity detection}
\acro{SBNR}{signal-to-babble-noise ratio}
\acro{SSNR}{signal-to-sensor-noise ratio}
\acro{SCM}{signal covariance matrix}
\acro{SDNR}{signal to diffuse noise ratio}
\acro{SH}{spherical harmonics}
\acro{SIR}{signal-to-interference ratio}
\acro{SMA}{spherical microphone array}
\acro{SMIRgen}{spherical microphone arrays impulse response
generator}
\acro{SPP}{speech presence probability}
\acro{SPEAR}{SPeech Enhancement for Augmented Reality}
\acro{STFT}{short-time Fourier transform}
\acro{STOI}{short-time objective intelligibility measure}
\acro{SFT}{spherical Fourier transform}
\acro{SH}{spherical harmonic}
\acro{SNR}{signal-to-noise ratio}
\acro{SDR}{signal-to-distortion ratio}
\acro{SAR}{signal-to-artifact ratio}
\acro{SWNR}{signal to spatially-white noise ratio}
\acro{SS}{Signal Subspace}
\acro{TF}{time-frequency}
\acro{ULA}{uniform linear array}
\acro{VAD}{voice activity detector}
\acro{WER}{word error rate}
\acro{WGN}{white Gaussian noise}
\acro{WSTOI}{weighted \acs{STOI}}

% multicols package not used
%\onecolumn  % add this line after \begin{document} but before \titlepage

\title{Subspace Hybrid MVDR Beamforming for Augmented Hearing}

% Authors file in here:
\author{%
Sina Hafezi, %
Alastair H. Moore, %
Pierre H. Guiraud, %
Patrick A. Naylor,~\IEEEmembership{Fellow,~IEEE}\\ %
Jacob Donley, %
Vladimir Tourbabin,~\IEEEmembership{Member,~IEEE}, %
Thomas Lunner
\thanks{S. Hafezi, A. H. Moore, P. Guiraud, and P. A. Naylor are with the Department of Electrical and Electronic Engineering, Imperial College London, UK (e-mail: \{sina.hafezi,~alastair.h.moore,~p.guiraud,~p.naylor\}@imperial.ac.uk).}%
\thanks{J. Donley, V. Tourbabin, and T. Lunner are with Meta Reality Labs Research, Redmond, Washington, USA (e-mail: \{jdonley,~vtourbabin,~thlu\}@meta.com).}%
\thanks{This work was supported by Meta Reality Labs Research.}%
}

\maketitle %
%
% multicols package not used
%\begin{multicols}{2}  %Before starting of \begin{abstract}

% UNCOMMENT THIS FOR GLOSSARY OF TERMS:
%\input{glossary.tex}

\begin{abstract}
Signal-dependent beamformers are advantageous over signal-independent beamformers when the acoustic scenario - be it real-world or simulated - is straightforward in terms of the number of sound sources, the ambient sound field and their dynamics. However, in the context of augmented reality audio using head-worn microphone arrays, the acoustic scenarios encountered are often far from straightforward. The design of robust, high-performance, adaptive beamformers for such scenarios is an on-going challenge. This is due to the violation of the typically required assumptions on the noise field caused by, for example, rapid variations resulting from complex acoustic environments, and/or rotations of the listener's head. This work proposes a multi-channel speech enhancement algorithm which utilises the adaptability of signal-dependent beamformers while still benefiting from the computational efficiency and robust performance of signal-independent super-directive beamformers. The algorithm has two stages. (i) The first stage is a hybrid beamformer based on a dictionary of weights corresponding to a set of noise field models. (ii) The second stage is a wide-band subspace post-filter to remove any artifacts resulting from (i). The algorithm is evaluated using both real-world recordings and simulations of a cocktail-party scenario. Noise suppression, intelligibility and speech quality results show a significant performance improvement by the proposed algorithm compared to the baseline super-directive beamformer. A data-driven implementation of the noise field dictionary is shown to provide more noise suppression, and similar speech intelligibility and quality, compared to a parametric dictionary.
\end{abstract}


% \fi
\IEEEpeerreviewmaketitle

% all the other chapters in here:
\acresetall
\section{Introduction}
\label{sec:introduction}
% \acresetall

Colin Cherry was the first to identify the `cocktail party problem' as the task of enhancing one target signal, or voice, in the presence of multiple interfering signals and noise \citep{Cherry1953,Bronkhorst2000}. Despite advances in speech enhancement and microphone array processing, the cocktail party problem still remains a challenge in the audio signal processing community after decades of research \citep{Bronkhorst2000,Donley2021a,SPEARwebsite}. Relevant applications include teleconferencing, speech recognition in human-machine interactions, hearing aids, and augmented reality audio. 

It is widely known that significant enhancement can be achieved using the acoustic signals captured with a microphone array device \citep{Doclo2010,Lollmann2017,Klasen2007,HaebUmbach2019}. The use of wearable microphone arrays is a beguiling approach towards addressing the cocktail party problem. Wearable arrays, such as \ac{AR} glasses \citep{Donley2021a}, can additionally provide head/array position and orientation information, as well as source \acp{DOA} and/or other environmental information obtained using a camera or other sensors. However, the compact size and real-world implementation of such systems restricts the processing to be computationally inexpensive, causal, real-time and robust to unforeseen scenarios. This paper addresses scenarios with one or more target talkers in the presence of localized interferers and ambient noise, using a head-worn microphone array mounted in \ac{AR} glasses. It is assumed in the following that the array's \acp{ATF} and the target \acp{DOA} are known with realistic accuracy.

Multichannel speech enhancement systems \citep{Benesty2008b} typically employ \ac{MISO} algorithms exploiting spatial, temporal and spectral information to improve the \ac{SNR} at the output. The potential for additional improvement using \ac{SISO} post-processing, such as the Wiener filter \citep{Wiener1949}, spectral subtraction \citep{Loizou2007} or \ac{DL}-based methods \citep{Hu2020a,Luo2018a} is recognized but not studied here; the focus in this work is the \ac{MISO} block in such systems. Without loss of generality, a multi-target scenario is considered as multiple single-target scenarios in which the same \ac{MISO} method can be applied. Such decomposition of the scenario is known to improve the enhancement performance due to the use of fewer distortionless target constraints and correspondingly increase the potential for noise suppression. In addition, it enables control of the mixture of enhanced sources in the output depending on the application.

In this work, a novel analytical beamforming algorithm is proposed which is causal and able to adapt to complex and non-isotropic noise fields while being computationally efficient due to the use of pre-calculated weights. The dictionary of precalculated beamformer weights is shown to have the flexibility of being analytical (parametric) or data-driven. Early work on this topic was described in \citep{Hafezi2023}. Here, the work has been extended to include a new formulation, the data-driven dictionary, more extensive evaluations and mathematical analysis of the method.

The remainder of the paper is structured as follows. Section~\ref{sec:backgroundReview} provides a brief review of beamforming approaches. Section~\ref{sec:problemformulation} formulates the problem and defines the baseline beamformer. Sections~\ref{sec:method} and \ref{sec:dictionary} respectively propose the multichannel enhancement system and two categories of parametric and data-driven dictionaries of beamformer weights. Section~\ref{sec:evaluation} compares the two implementations of the proposed method with the baseline beamformer using real-world recordings and simulated cocktail party data with a head-worn microphone array. Finally, conclusions are given in Section~\ref{sec:conclusion}.

\section{Background Review}
\label{sec:backgroundReview}

Beamforming algorithms spatially filter a sound field with the aim of preserving signals from the target direction(s) while minimizing the power of all other signals in the output. 

\subsection{Classical Beamforming Algorithms}

The implementation of classical beamformers is characterized by a filter-and-sum structure, either in the time domain or \ac{STFT} domain, and is of relatively low computational complexity \citep{Gannot2017}. Signal-independent beamformers, such as delay-and-sum \citep{Dudgeon1984} and fixed versions of \ac{MVDR} \citep{Capon1969} use static, possibly switched, precalculated filter coefficients. On the other hand, adaptive beamformers such as \ac{MVDR} or \ac{LCMV} \citep{Trees2002a} with dynamic covariance input can potentially improve performance by adapting their spatial filtering characteristics to a particular, and possibly time-varying, acoustic scenario.  Such signal-dependent beamformers rely on the use of a \ac{SCM} or \ac{NCM}, resulting in the  \ac{MPDR}\footnote{We note that beamformer nomenclature varies and that the term \acs{MVDR} is used by some authors for either or both of the \acs{MPDR} and \acs{MVDR} beamformers.} \citep{Trees2002_6,Capon1969} or  \ac{MVDR} beamformers, respectively. The \ac{MVDR} approach is often preferred due to real-world inaccuracies in the steering vector causing potential performance degradation in \ac{MPDR} because of target signal cancellation \citep{Cox1973a,Ehrenberg2010a}. 

A common and robust choice for the \ac{NCM} in \ac{MVDR} implementations is a signal-independent model. Examples include the matched-filter beamformer \citep{Jan1996,Doclo2010} in which the \ac{NCM} is a multiple of the identity matrix, and the superdirective beamformer \citep{Bitzer2001a} in which the \ac{NCM} is the covariance matrix corresponding to a spherically or cylindrically isotropic diffuse noise field. The use of a signal-dependent \ac{NCM} potentially improves spatial filtering performance if it is estimated accurately. Some such adaptive beamformers aim to estimate \ac{NCM} either directly during periods when the target signal is inactive, or indirectly by subtracting the target contribution from the \ac{SCM}. Both these approaches require identification of \ac{TF} bins containing target signal activity using, for example, \ac{SAD} or \ac{SPP} methods \citep{Gannot2017}. 

In \citep{Moore2022a}, the \ac{SCM} is decomposed into isotropic, identity and \ac{PW} components followed by removal of any (near-) target \acp{PW} from the modelled \ac{SCM} to avoid signal cancellation. Although some signal-dependent beamformers are shown \citep{Moore2022a} to perform better than signal-independent beamformers, such evaluations are typically limited to simulations or simple-scene real-recordings. These typically avoid the real-world complexities of the cocktail party problem with wearable arrays such as rapid dynamics of the noise field due to head rotation or the presence of non-isotropic noise fields.

\subsection{Deep Learning Approaches}
Typical \ac{DL}-based \ac{MISO} speech enhancement methods use neural networks to estimate either the output (end-to-end methods) \citep{Pandey2022a}, or other information such as the \ac{NCM} or beamformer weights (`neural beamforming') \citep{casebeer2021nicebeam}. In \citep{Heymann2016a,Wang2018b,Wang2020,Zhang2021}, real- or complex-valued \ac{TF} ratio masks are estimated while in \citep{Liu2018,Erdogan2016} the \ac{TF} bins dominated by noise are identified to estimate the \ac{NCM} for use in \ac{MVDR} beamforming. The work in \citep{Ren2021} employs a U-Net architecture to directly estimate complex-valued beamforming weights. An extension of this work using DCCRN \citep{Hu2020a} is also proposed in \citep{Chen2022a}. Similarly to other neural beamformers, FaSNet \citep{Luo2019a} aims to estimate the beamformer filters but in the time-domain using a temporal convolutional network. Although the evaluation results from some of these methods are exceptionally good, these systems are limited to offline and/or high-latency applications due to non-causality and relatively heavy computational load. Methods such as \citep{Ren2021,Luo2019a} are causal and designed for low-latency purposes. However, so far, they have not been evaluated in real-world, unseen cocktail party scenarios. In addition, \ac{DL}-based methods are typically trained for only a particular architecture, input format, sample rate, and array geometry. In order to handle extra inputs, such as head tracking information or changes in the above configurations, the models usually require re-training and, in some cases, re-designing.
% add sections as required
\section{Formulation and Baseline Beamformer}
\label{sec:problemformulation}
%\acresetall

For notational simplicity, the formulation is provided for a single-channel output. However, the same procedure can be repeated assuming different reference channels for multi-channel output. 

\subsection{Signals in the \ac{STFT} domain}

Let $\stftArray_{\iMic}(\TF)\in\Complex$ denote the signal at microphone index $\iMic$, frequency index $\iFreq$ and time-frame index $\iFrame$. Using a \ac{PW} representation, this can be expressed as  
\begin{equation}
\stftArray_{\iMic}(\TF) = \sum_{\iPW=1}^{\nPW} \stftPW_{\iMic,\iPW}(\TF) 
+ \stftWhite_{\iMic}(\TF),
\label{eq:PWD}
\end{equation}
where $\stftPW_{\iMic,\iPW}\in\Complex$ and $\stftWhite_{\iMic}\in\Complex$ are, respectively, the $\iPW$-th plane-waves with unique \acp{DOA}, and the sensor noise at microphone $\iMic$. For brevity, the $(\TF)$ dependency will be omitted in the following when unambiguous.

Defining $\stftArrayVec = [\stftArray_{1}\ldots\stftArray_{\nMic}]\transposeOp\in\Complex^{\nMic\times1}$ as the vector of noisy microphone  signals, the beamformer output is
\begin{equation}
    \stftOut = \stftWeightVec\hermitianOp\stftArrayVec,\label{eq:beamforming}
\end{equation}
where $\stftWeightVec\in\Complex^{\nMic\times1}$ is the vector of beamformer weights and $(.)\hermitianOp$ indicates Hermitian transpose. 

\subsection{MVDR}

In the \ac{MVDR} beamformer design, the weights are given by \citep{Capon1969}
\begin{align}
\stftWeightVec & =\frac{\covRobust^{-1}\steerVec}{\steerVec^{H}\covRobust^{-1}\steerVec}%
\label{eq:MVDRweights}\\
\mathrm{with}\quad\covRobust & =\covNCM+\scalarLoading\eye\label{eq:robustCovMat}\\
\mathrm{and}\quad\scalarLoading & =\max\left(\frac{\covEigenvalue_\mathrm{max}-\covCondNumLimit\covEigenvalue_\mathrm{min}}{\covCondNumLimit-1},\,0\right),\label{eq:robustEpsilon}
\end{align}
where $\steerVec = \tilde{\ATF}(\direction\Target)\in\Complex^{\nMic\times1}$ is the steering vector for the target \ac{DOA} $\direction\Target$, and $\tilde{\ATF}(\cdot)$ is the array's \ac{RTF} that is determined as the array's \ac{ATF} $\ATF(\cdot)$ normalized by the reference output channel. The $\covNCM\in\Complex^{\nMic\times\nMic}$ is an estimate of 
the \acs{NCM}, $\eye\in\Complex^{\nMic\times\nMic}$ is the identity matrix, $\covEigenvalue_\mathrm{max}$ and $\covEigenvalue_\mathrm{min}$ are the largest and smallest eigenvalues of $\covNCM$, and $\covCondNumLimit$ is the maximum permitted condition number of $\covRobust$ and, indirectly, constrains the norm of $\stftWeightVec$ and the sensitivity of the beamformer to errors in $\steerVec$ \citep{Cox1987,Li2003}.

\subsection{Isotropic-\ac{MVDR} (Superdirective) Beamformer} 
\label{ssection:superdirective}

The baseline beamformer in this study is chosen to be the Isotropic-\ac{MVDR} referred to as `Iso' (also known as Maximum Directivity), in which a static, spherically diffuse sound field is assumed for the $\covNCM$ in \eqref{eq:robustCovMat}. This is equivalent to assuming the presence of uncorrelated interference with equal power from all directions. The spherically isotropic diffuse covariance matrix can be obtained as

\begin{equation}
	\covDiffuse = \int_{\direction} \ATF(\direction) {\ATF\hermitianOp}(\direction) d\direction, 
\label{eq:covdiffuse}
\end{equation}
where $\int_{\direction} d\direction= \int_{0}^{2\pi} \int_{0}^{\pi} \sin(\inc) d\inc d\az$ denotes integration along azimuth $\az\in[0,2\pi)$ and inclination $\inc\in[0,\pi]$.

In the case of using a discrete set of $\direction\in\DOAset$ from a uniform grid of directions, \eqref{eq:covdiffuse} can be approximated by quadrature weighting giving

\begin{equation}
   \covEstIso = \sum_{i \in \DOAset} \quadWeight_{i}\ATF(\direction_{i}) {\ATF\hermitianOp}(\direction_{i}),
\label{eq:Riso}
\end{equation}
where $\quadWeight_{i}\in\Real$ is the quadrature weight for each sample direction of $\ATF$ given by \citep{Driscoll1994} as
\begin{equation}
\quadWeight_{i} = \frac{2\sin \inc_i}{N_\az N_\inc}\sum_{p=0}^{\lceil 0.5N_\inc-1 \rceil}\frac{\sin\left(\left(2p+1\right)\inc_i\right)}{2p+1},
\label{eq:quadWeights}
\end{equation}
in which $\inc_i$ is the inclination of sample point $i$ and  the number of sample points in azimuth and inclination are $N_\az$ and $N_\inc$ respectively. 

Note that the quadrature weighting is done to preserve the uniform power isotropy by compensating for the higher density of points closer to the poles in a uniform grid spatial sampling scheme. The use of other spatial sampling schemes may require no or different weighting. Using $\covNCM=\covEstIso$ in \eqref{eq:robustCovMat} and substituting into \eqref{eq:beamforming}, the output of this beamformer is denoted as $\stftOut\Iso$.
\section{Proposed Method}
\label{sec:method}

\begin{figure}[t]
    \centering
    \includegraphics[scale=.4]{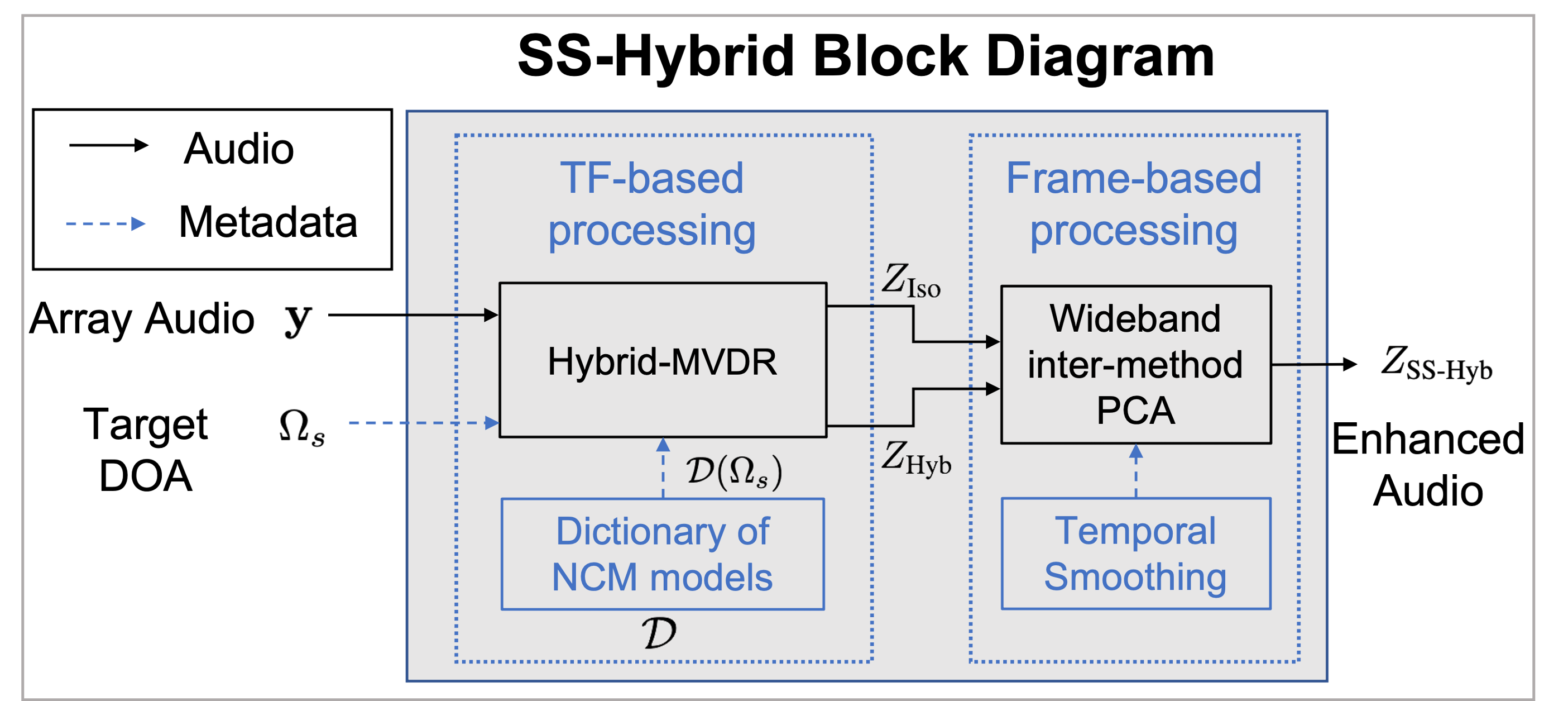}
    \caption{The proposed system block diagram.}
    \label{fig:block}
\end{figure}

As illustrated in Fig. \ref{fig:block}, the proposed method consists of two stages. In the first stage (Hybrid-\ac{MVDR}), multiple \ac{MVDR} beamformers are implemented, each with a different precalculated $\covNCM$, including isotropic and non-isotropic noise field models. The first stage outputs two signals: $\stftOut\Iso$ being the result of Iso and $\stftOut\Hyb$ being the output from the Hybrid-\ac{MVDR} beamforming. The second stage applies wideband inter-method \ac{PCA}. This processes the two full-band spectral outputs from the first stage in order to remove the `musical noise' generated in the output of the Hybrid-\ac{MVDR}. 

\subsection{Hybrid-MVDR}

Multiple \ac{MVDR} beamformers are implemented with the same steering vector but various \acp{NCM} taken from a dictionary of predefined and precalculated noise field models. The dictionary is denoted 

\begin{equation}
    \dict=\{\covMat_{\iModel,\iFreq} \rightarrow \stftWeightVec_{\iModel,\iFreq,\direction}\}\quad\text{for}\quad\forall\iFreq\quad\text{and}\quad\direction\in\SteeringSet \label{eq:dictionary}
\end{equation}
and contains the pre-calculated beamformer weights $\stftWeightVec_{\iModel,\iFreq,\direction}\in\Complex^{\nMic\times1}$. The weights are derived using \eqref{eq:MVDRweights} for $\iModel=1, \dots, \nModel$ \ac{NCM} models, $\covRobust$, and a discrete set $\SteeringSet$ of possible steering directions $\direction$. Note that the dictionary $\dict\in\Complex^{\nModel\times\nFreq\times\nMic\times\nDir}$ is a $4$-dimensional table where $\nModel$, $\nFreq$, $\nMic$ and $\nDir$ are respectively the total number of \ac{NCM} models, frequency bands, microphones and steering directions.

The detailed choices of the \ac{NCM} models in the dictionary $\dict$ will be discussed in Section \ref{sec:dictionary}. For simplicity and as a necessity for the description of this section, it should be noted that the dictionary $\dict$ includes the isotropic model, $\covEstIso$ in \eqref{eq:Riso}, as well as other models to be specified later.

At each \ac{TF} bin $(\TF)$, and for a known steering direction $\direction\Target(\iFrame)$, the Hybrid-\ac{MVDR} in Fig. \ref{fig:block} implements multiple \ac{MVDR} beamformers using precalculated weights associated with the closest steering direction $\direction\Target(\iFrame)$ stored in $\dict$. The output with the minimum power among the models is selected and denoted as 
\begin{align}
    & \stftOut\Hyb= (\stftWeightVec_{j})^{H}\stftArrayVec,\label{eq:hyb}\\
    \mathrm{with\quad} & j = \argmin_{m}{\{\lVert (\stftWeightVec_{m})^{H}\stftArrayVec \rVert^{2}\}}\quad\forall \stftWeightVec_{m}\in\dict(\direction\Target).\label{eq:minimisation}
\end{align}
The isotropic model is included in $\dict$ and the corresponding beamformer $\stftOut\Iso$ is output to the next stage as shown in Fig.~\ref{fig:block}.

As will be shown in Section~\ref{sec:evaluation}, although the Hybrid beamformer typically results in stronger acoustic noise reduction compared to Iso, the output may contain `musical noise' due to rapid switching of beam pattern between neighbouring time-frames and frequencies. To suppress this musical noise, which is assumed to be sufficiently uncorrelated with the acoustic noise, the second stage extracts the component of the Hybrid-\ac{MVDR} output which is correlated with the Iso-\ac{MVDR} output. 

\subsection{Wideband inter-method PCA}
\label{ssec:pca}
In this stage, a solution is proposed for the removal of the musical noise in the $\stftOut\Hyb$ while preserving its stronger noise suppression compared to $\stftOut\Iso$. This section formulates the proposed solution whereas the Appendix provides a corresponding theoretical analysis.

Defining $\check{(.)}$ as the vector of full-band spectrum, let $\check{\stftOutVec}\Iso(\iFrame)=[\stftOut\Iso(1,\iFrame),\dots,\stftOut\Iso(\nFreq,\iFrame)]\in\Complex^{1\times\nFreq}$ and $\check{\stftOutVec}\Hyb(\iFrame)=[\stftOut\Hyb(1,\iFrame),\dots,\stftOut\Hyb(\nFreq,\iFrame)]\in\Complex^{1\times\nFreq}$ denote the full-band spectral outputs of Iso and Hybrid-\ac{MVDR} beamformers at time-frame $\iFrame$, which are then concatenated to form a two-channel wideband data matrix $\stftOutMat\in\Complex^{2\times\nFreq}$  defined as

\begin{equation}
    \stftOutMat(\iFrame) = \begin{bmatrix}
    \check{\stftOutVec}\Hyb(\iFrame)\\
    \check{\stftOutVec}\Iso(\iFrame)\end{bmatrix}.\label{eq:MethodsArray}
\end{equation}
The inter-method wideband covariance matrix $\covMat\Out(\iFrame)\in\Complex^{2\times2}$ is then defined as

\begin{equation}
    \covMat\Out(\iFrame)=\expectOp\{\stftOutMat(\iFrame)\stftOutMat(\iFrame)\hermitianOp\},\label{eq:MethodsCovMat}
\end{equation}
where  $\expectOp\{.\}$ is the expectation operator. An estimate of $\covMat\Out$ in \eqref{eq:MethodsCovMat} can be obtained by applying an \ac{EMA} process to the instantaneous inter-method covariance matrix as

\begin{equation}
    \hat{\covMat}\Out(\iFrame) = \smoothness\hat{\covMat}\Out(\iFrame-1) + (1-\smoothness)\stftOutMat(\iFrame)\stftOutMat(\iFrame)\hermitianOp,\label{eq:MethodsCovMatEst}
\end{equation}
where $0<\smoothness=e^{-\Delta t / \timeConst}<1$ is the smoothing factor, $\Delta t$ is the \ac{STFT} time-frame step, and $\timeConst$ is a forgetting time constant.

Using \ac{EVD}, 
\begin{equation}
    \hat{\covMat}\Out(\iFrame) = \EigenVec(\iFrame)\EigenVal(\iFrame)\EigenVec^{-1}(\iFrame),\label{eq:EVD}
\end{equation}
where $\EigenVec\in\Complex^{2\times2}$ and $\EigenVal\in\Real^{2\times2}$ are, respectively, the eigenvectors and diagonal matrix of eigenvalues. Assuming the columns of $\EigenVal$ are sorted in descending order of eigenvalues, the first column of $\EigenVec(\iFrame)=[\EigenVec_{S}(\iFrame),\EigenVec_{N}(\iFrame)]$ is considered to be the signal eigenvector denoted as $\EigenVec_{S}(\iFrame)\in\Complex^{2\times1}$.

The \ac{SS} of the $\stftOutMat(\iFrame)$ is reconstructed as 
\begin{equation}
    \stftOutMat\SubS(\iFrame)=\EigenVec_{S}(\iFrame)\EigenVec_{S}\hermitianOp(\iFrame)\stftOutMat(\iFrame),\label{eq:SS}
\end{equation}
where the first row of $\stftOutMat\SubS(\iFrame)=[\check{\stftOutVec}\SSHyb(\iFrame)\transposeOp,\check{\stftOutVec}\SSIso(\iFrame)\transposeOp]\transposeOp\in\Complex^{2\times\nFreq}$ is the subspace-Hybrid (SS-Hybrid) output. In a compact form, the final output spectrum can be written in terms of weighted mixing of $\check{\stftOutVec}\Iso$ and $\check{\stftOutVec}\Hyb$ spectra as
\begin{align}
    & \check{\stftOutVec}\SSHyb(\iFrame) = \beta\Hyb(\iFrame)\check{\stftOutVec}\Hyb(\iFrame) + \beta\Iso(\iFrame)\check{\stftOutVec}\Iso(\iFrame),\label{eq:SSHyb} \\
    & \beta\Hyb(\iFrame) = \EigenVecVal_{1}\EigenVecVal^{*}_{1},\label{eq:betaHyb} \\
    & \beta\Iso(\iFrame) = \EigenVecVal_{1}\EigenVecVal^{*}_{2},\label{eq:betaIso}
\end{align}
where the $\EigenVecVal_{1}$ and $\EigenVecVal_{2}$ are respectively the first and the second element in the subspace eigenvector $\EigenVec_{S}(\iFrame)=[\EigenVecVal_{1},\EigenVecVal_{2}]\transposeOp$ and $(\cdot)^{*}$ denotes the complex conjugate. A mathematical analysis of $\stftOutMat$ and $\hat{\covMat}\Out$ in terms of the target and residual noise from each beamformer is given in the Appendix.

\section{Dictionary}
\label{sec:dictionary}

This section proposes two alternative dictionary designs $\dict$ for the Hybrid-\ac{MVDR}: (A) Parametric or (B) Data-driven. The parametric dictionary consists of \ac{NCM} models $\covMat$ that are derived using array \acp{ATF} and a set of pre-defined noise field power distributions. The data-driven dictionary includes \ac{NCM} models derived from a set of array recordings assuming the ground truth noise recordings are available. Details of these two categories are discussed next.

\begin{figure}[b]
    \centering
    \includegraphics[scale=.21]{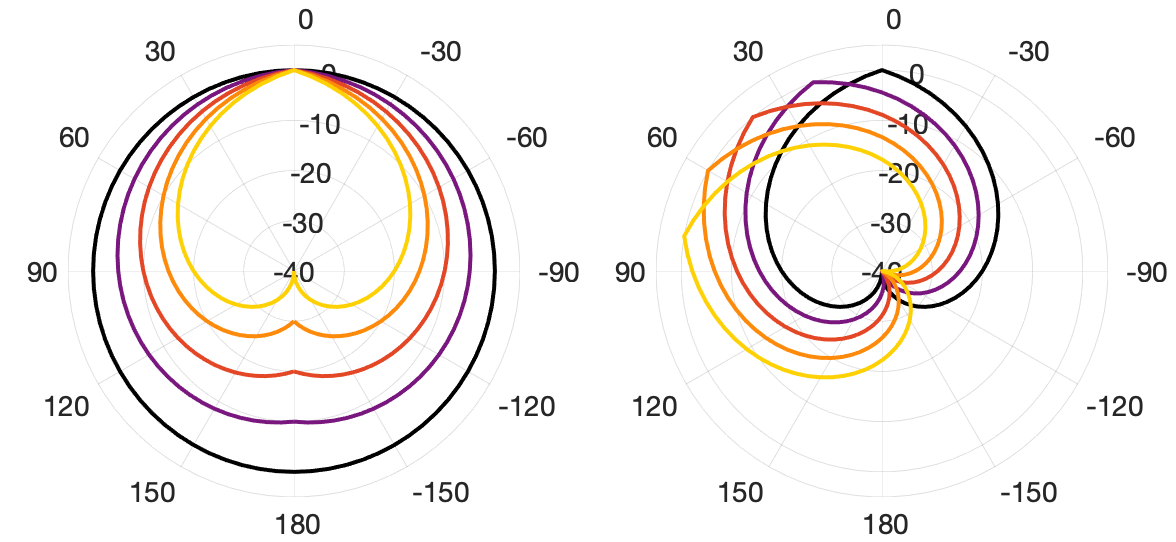}
    \caption{(Left) Isotropic and horizontally unimodal Anisotropic models with a fixed $\az\uPeak=0\deg$ and varying $\powDyn=\{0, 10, 20, 30, 40\}$\,dB. (Right) Anisotropic models with varying $\az\uPeak=\{0, 20,40, 60, 80\}\deg$ and fixed $\powDyn=40$\,dB.}
    \label{fig:aniso}
\end{figure}

\subsection{Parametric Dictionary}

A parametric dictionary can be formed by considering both spatially correlated  and uncorrelated \ac{NCM} models. An example of a spatially uncorrelated noise model is sensor noise that has no inter-channel correlation, whereas an example of a spatially correlated noise model is environmental noise with arbitrary isotropy. 

It can be noted that, for spatially uncorrelated models, the \ac{NCM}, $\covMat$, is directly formulated whereas for the spatially correlated models, the formulation is done for the noise field isotropy $\isotropy(\direction)$, which then can be used along with the available \acp{ATF} to obtain the \ac{NCM} model $\covMat$ as
\begin{equation}
    \covMat = \sum_{i \in \DOAset} \isotropy(\direction_{i})\quadWeight_{i}\ATF(\direction_{i}) {\ATF\hermitianOp}(\direction_{i}).
\label{eq:isotropy2NCM}
\end{equation}

\subsubsection{Identity (Spatially uncorrelated model)}
\label{sssection:model_Identity}

This model results in a diagonal \ac{NCM} where the diagonal values represent the relative noise power at each sensor. Assuming a spatially and spectrally uniform (white) sensor noise power, this model is given as 
 \begin{equation}
     \covMat\Identity = \eye,\label{eq:dict_Identity}
 \end{equation}
where $\eye\in\Complex^{\nMic\times\nMic}$ is the identity matrix. Note that the identity matrix can be replaced with a diagonal matrix with non-uniform values if the array sensors' noise powers are available, non-uniform and spectrally non-white.

\begin{figure*}[t]
    \centering
    \includegraphics[scale=.49]{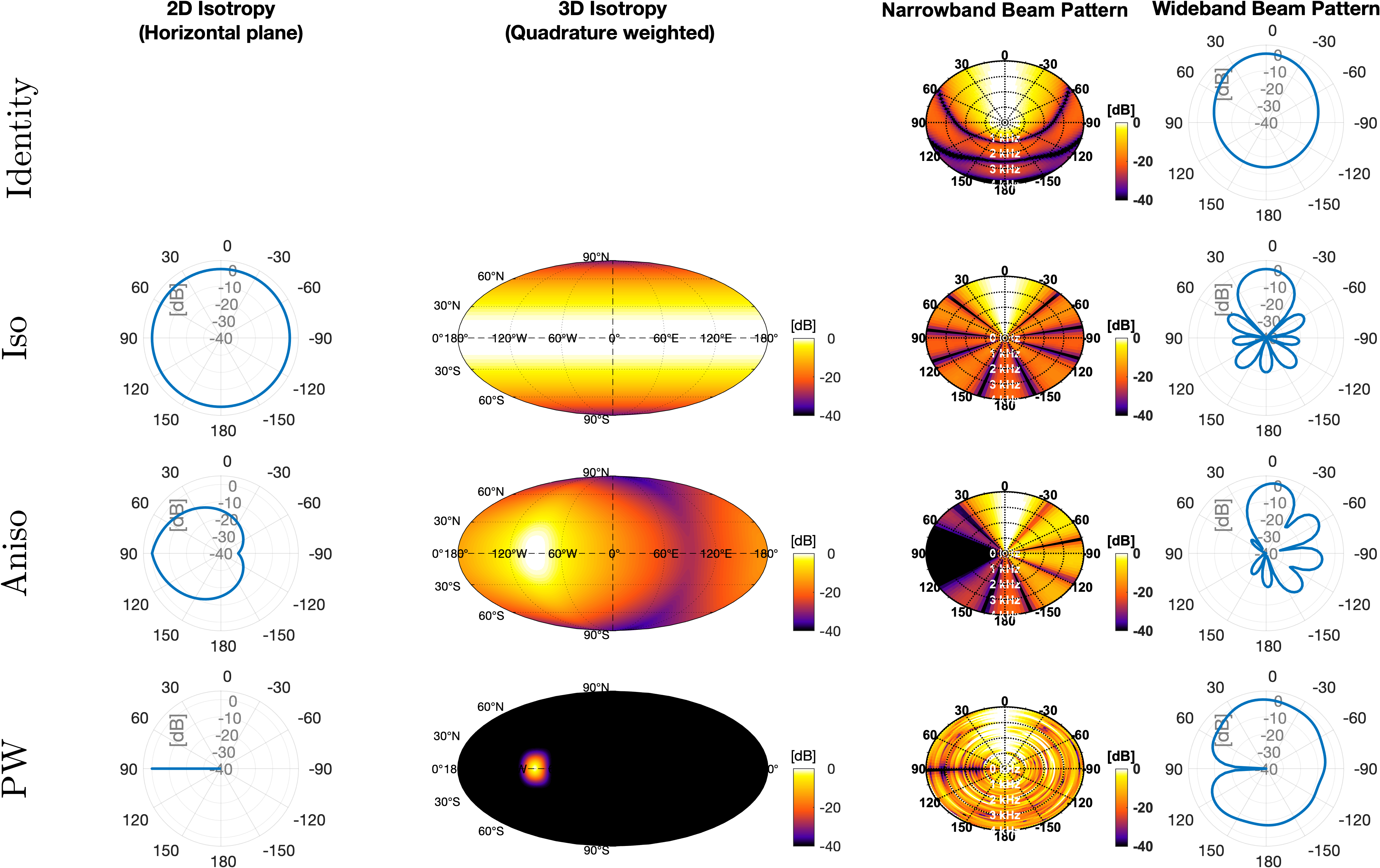}
    \caption{Rows: Identity, Iso, unimodal Aniso and PW models with $\az\uPeak=90\deg$. Columns: \ac{NCM} 2D isotropy (horizontal plane), \ac{NCM} 3D power distribution (quadrature weighted along the inclination), narrowband and wideband (a-weighted) beam patterns (horizontal plane) with steering direction $\direction\Target=(\az\Target,\inc\Target)=(0,90)\deg$.}
    \label{fig:models}
\end{figure*}

\subsubsection{Isotropic model}
\label{sssection:model_iso}

The spherically isotropic model has uniform isotropy across all directions and is given as 
\begin{equation}
    \isotropy\Iso(\direction) = 1.\label{eq:dict_iso}
\end{equation}
Substituting \eqref{eq:dict_iso} into \eqref{eq:isotropy2NCM} results in $\covMat\Iso$, which is equivalent to the $\covMat$ used in \eqref{eq:Riso} for the superdirective beamformer in Section~\ref{ssection:superdirective}. 

\subsubsection{Anisotropic models}
\label{sssection:model_aniso}

Anisotropic models can be defined in infinitely many ways. However, in this work, only horizontally unimodal (single-peak) isotropy is considered. Assuming a linear function for horizontal unimodality, the anisotropic model with a mode azimuth $\az\uPeak$ and power dynamic range $\powDyn$ (in decibels) is given as
\begin{equation}
    \isotropy\Aniso(\direction) = 10^{-\powDyn|\angle(\az,\az\uPeak)|/(10\pi)},\label{eq:dict_aniso}
\end{equation}
where $\angle(\cdot\,,\cdot)$ is the angular difference in radians between two directions wrapped to $[-\pi,\pi]$.
Note that the linearity of isotropy is on the logarithmic scale for the power, independent of the inclination $\inc$ and symmetric around the mode azimuth $\az\uPeak$. The $\covMat\Aniso$ can be obtained by substituting \eqref{eq:dict_aniso} into \eqref{eq:isotropy2NCM}. Figure~\ref{fig:aniso} illustrates the horizontally unimodal anisotropic noise field models for a few examples with fixed $\az\uPeak$ and varying $\powDyn$ (left), as well as varying $\az\uPeak$ and fixed $\powDyn$ (right).

\subsubsection{Plane-wave models}
\label{sssection:model_pw}

The \ac{PW} model represents the effect of a single \ac{PW} with \ac{DOA} $\direction\uPeak$ in matrix $\covMat$. This is formulated as
\begin{align}
    \isotropy\PW(\direction) & =\begin{cases}
1 & \mathrm{for}\quad\direction=\direction\uPeak\\
0 & \mathrm{otherwise}
    \end{cases}.\label{eq:dict_pw}
\end{align}
Empirical analysis suggested the removal of \ac{PW} models from our implementations and evaluations. They were shown to be the most frequently chosen model in the Hybrid-\ac{MVDR}. This commonly resulted in a failure to suppress the reverberation during target activity, due to the corresponding beam-pattern caused by the extreme directionality of \ac{PW} model, as also shown in Fig.~\ref{fig:models}. An evaluation using \ac{PW} models in the dictionary is presented in \citep{Hafezi2023}.

Figure~\ref{fig:models} shows examples for the 1)~Identity, 2)~Iso, 3)~unimodal Aniso and 4)~\ac{PW} models (with $\az\uPeak=90\deg$ for Aniso and \ac{PW}) with their assumed 2D and 3D power distributions of \ac{NCM}. Additionally shown are the resulting narrowband and wideband beam-patterns with steering direction $\direction\Target=(\az\Target,\inc\Target)=(0,90)\deg$. These results were obtained for a 32-element rigid \ac{SMA} with a radius of $4.2$cm (corresponding to the em32
Eigenmike\textregistered \  \citep{MHAcoustics2013}), with a sampling rate of $\SI{8}{kHz}$ and uniform spatial grid sampling of $3\deg$ resolution across both azimuth and inclination for the \acp{ATF}.

\subsection{Data-driven Dictionary}

The data-driven dictionary contains \ac{NCM} models estimated from noise-only array signals. The noise-only signals can be obtained either from signal segments during which the target  is inactive, and/or exploiting ground truth noise signals if/when available. The term `ground truth noise signals' refers to the array signal excluding the anechoic (direct-path) target. Hence, it includes the sensor noise, ambient and interferer(s) noise, as well as the target reverberation, since an ideal beamformer is expected to suppress the target reverberation as well as other sources of noise. Some ground truth noise signals are assumed to be available and can be obtained by simulating noisy scenarios using array's \acp{ATF}.

Using \ac{EMA} on the instantaneous ground truth covariance matrix, an estimate of the ground truth \ac{NCM} can be derived as

\begin{multline}
	\covEstOracle(\TF) = \timeconst \covEstOracle(\iFreq,\iFrame-1) + \\ (1-\timeconst) (\stftArrayVec\Oracle(\TF) \stftArrayVec\Oracle^{\hermitianOp}(\TF)),
	\label{eq:NCM_oracle}
\end{multline}
where $\stftArrayVec\Oracle=\stftArrayVec-\stftArrayVec_{\mathrm{target}}$ and $\stftArrayVec_{\mathrm{target}}\in\Complex^{\nMic\times1}$ is the vector of anechoic target signals at the array sensors.

The number of the models $\nModel$ has no upper bound, unlike the number of channels $\nMic$, frequency bands $\nFreq$ and the steering directions $\nDir$ that are either defined in a wrapped space or imposed by the choice of the array and the \ac{STFT} settings. In order to restrict the number of models $\nModel$ without imposing any restriction on the number of ground truth \ac{NCM} observations, the set of measured $\covMat\Oracle$ are compressed into a fixed number of bases $\Bar{\covMat}$ via k-means clustering, as next described.

\subsubsection{Vectorization of \ac{NCM}}
\label{sssec:vectorization}

Since k-means clustering operates on real-valued data, the complex-valued \ac{NCM} matrix must be vectorized into real-valued vectors. The \ac{NCM} matrix $\covMat\in\Complex^{\nMic\times\nMic}$ has symmetric real values and conjugate complex values around its diagonal. In addition, the diagonal values are always real. This structure is utilized to form an efficient mapping between complex-valued matrix form denoted as $\covMat\in\Complex^{\nMic\times\nMic}$ and real-valued vector form denoted as $\covVec\in\Real^{\nMic^{2}\times1}$ for \ac{NCM}. 

\begin{figure}[t]
    \centering
    \includegraphics[scale=.4]{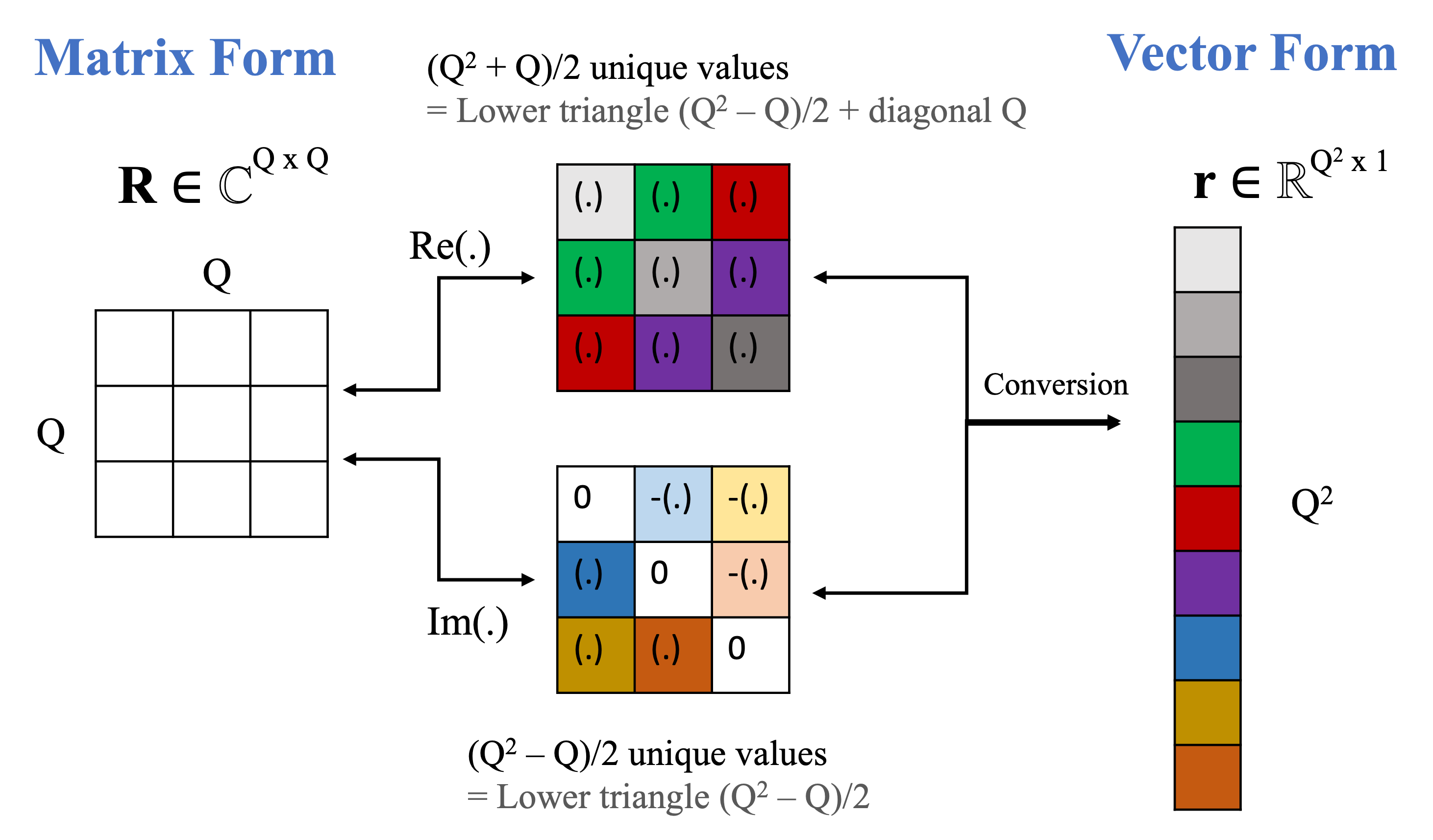}
    \caption{The process of conversion between complex-valued matrix form $\covMat$ and real-valued vector form $\covVec$ of \ac{NCM}.}
    \label{fig:vectorization}
\end{figure}

Figure~\ref{fig:vectorization} shows the process of conversion between the complex-valued matrix and real-valued vector forms. The complex-valued matrix $\covMat$ is split into real and imaginary parts followed by the concatenation of $\nMic^{2}$ unique real values from both real and imaginary matrices into one real-valued vector $\covVec$. This process is perfectly reversible.

\subsubsection{Clustering}
\label{sssec:clustering}

Having converted the arbitrary-size ($\nData$) set of estimated ground truth \acp{NCM} $\covRSet_{\iFreq} = \{\covMat\Oracle(\TF)\}\in\Complex^{\nData\times\nMic\times\nMic}$ per frequency band $\iFreq$ into its vectorized form $\covrSet_{\iFreq} = \{\covVec\Oracle(\TF)\}\in\Real^{\nData\times\nMic^{2}}$, k-means is then used to cluster $\covrSet_{\iFreq}\in\Real^{\nData\times\nMic^{2}}$ into $\Bar{\covrSet}_\iFreq = \{\Bar{\covVec}\Oracle(\iFreq,\iModel)\}\in\Real^{\nModel\times\nMic^{2}}$ centroids, where the number of clusters $\nCluster=\nModel$ is set to the total number of models $\nModel$ in the dictionary. The real-valued vectorized basis \ac{NCM} models are then converted into complex-valued matrix basis \acp{NCM} $\Bar{\covRSet}_{\iFreq} = \{\Bar{\covMat}\Oracle(\iFreq,\iModel)\}\in\Complex^{\nModel\times\nMic\times\nMic}$, which can be used in \eqref{eq:MVDRweights} to form the data-driven dictionary of beamformer weights. 

Note that the use of k-means clustering results in the number of models being independent of the training set size $\nData$. This enables the use of a compact dictionary with relatively small $\nModel$ for computational efficiency, while benefiting from training on a large set of signals with relatively large $\nData$. 
\section{Evaluations}
\label{sec:evaluation}

The algorithm was evaluated in the context of augmented hearing and augmented reality audio. Two implementations of the proposed method with parametric and data-driven dictionaries were compared with the Iso-\ac{MVDR} baseline beamformer, as well as the `passthrough' signal at the reference microphone. Data was obtained using a head-worn microphone array in real-recordings, and from simulated `cocktail party' scenarios.

\begin{figure}[b]
	\centering
	\centerline{\includegraphics[height=3.0cm]{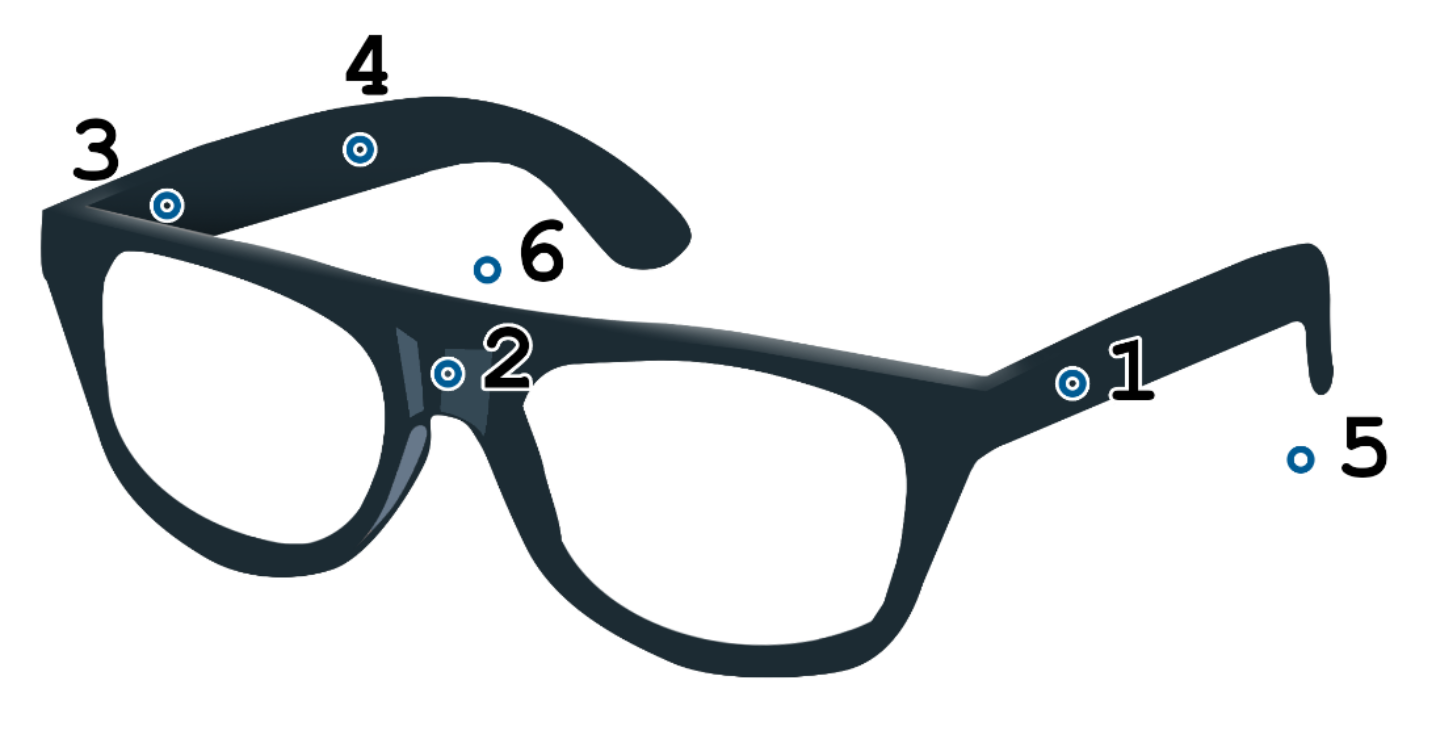}}
	\caption{Head-worn `glasses' microphone array used in EasyCom \citep{Donley2021a}. Microphones 5 and 6 are in-ear microphones.}
	\label{fig:EasyCom_array}
\end{figure}
\subsection{Head-worn microphone array}
\label{ssec:array}
The head-worn `glasses' microphone array from EasyCom dataset \citep{Donley2021a} was used. As shown in Fig.~\ref{fig:EasyCom_array}, it is a $6$-channel array with four microphones on the glasses and two in-ear microphones. The \acp{ATF} were measured for a quasi-uniform grid of directions with coverage range of $[-174\deg,180\deg]$ and $[-66\deg,84\deg]$ and resolution of $6\deg$ and $9\deg$ across azimuth and elevation, respectively. The measurement was done using a manikin head wearing the array in an anechoic chamber of Meta Reality Labs.

\subsection{Dataset}

\begin{figure}[b]
	\centering
	\centerline{\includegraphics[height=5cm]{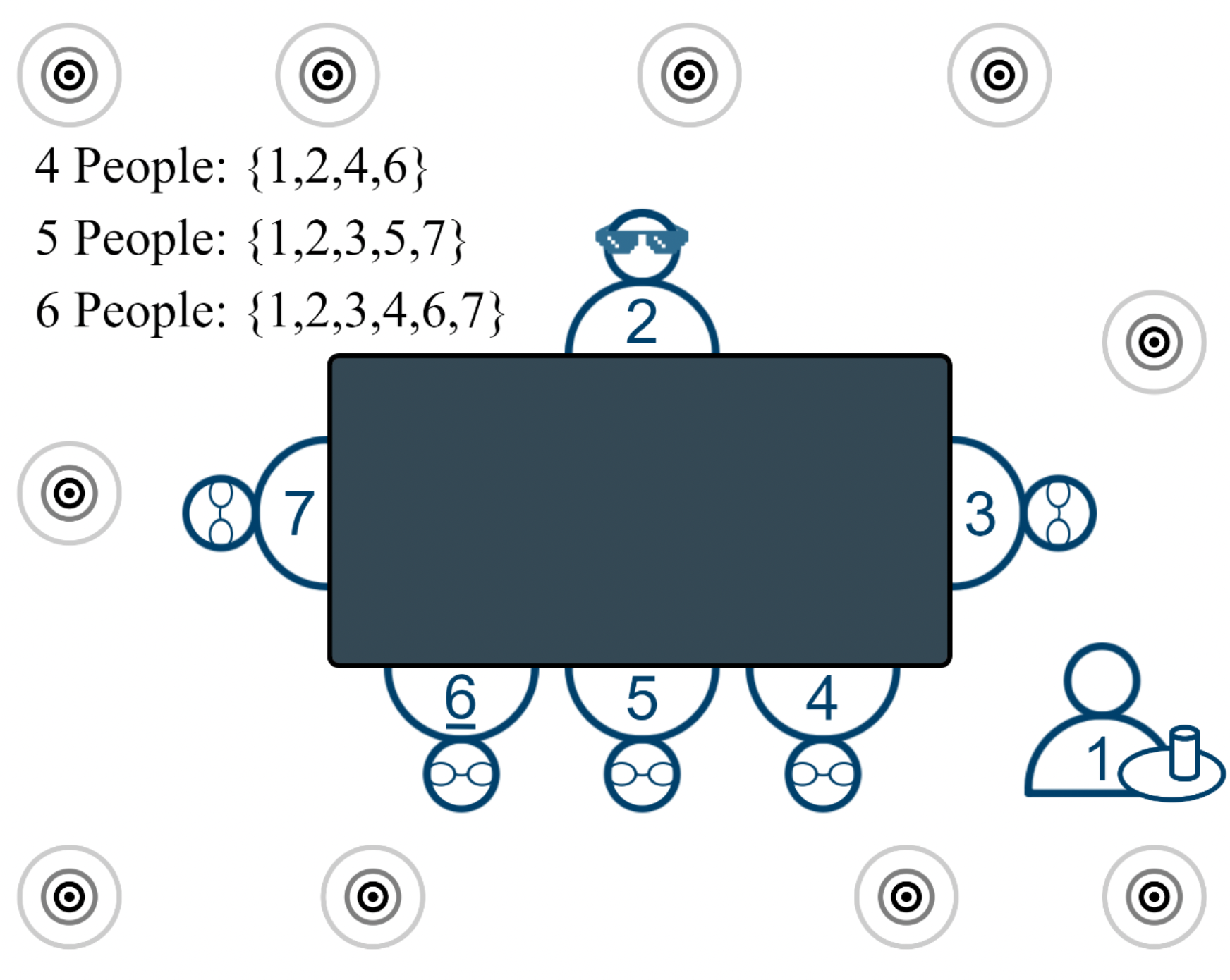}}
	\caption{Schematic of the scene used in EasyCom \citep{Donley2021a}. The `waiter', ID=1, is not wearing a close-talking microphone and their position is not tracked.
	}
	\label{fig:EasyCom_scene}
\end{figure}

The evaluation was performed using part of the \ac{SPEAR} dataset \citep{SPEARwebsite}, which is an extension of the EasyCom dataset \citep{Donley2021a}. The EasyCom dataset contains several realizations of a scenario illustrated in Fig.~\ref{fig:EasyCom_scene} in which a small group of people, sitting around a table in a room, have a natural conversation surrounded by ten loudspeakers playing restaurant-like ambient noise. The \ac{SPEAR} dataset is split into sub-datasets where the first, D1, is the entire real-recording EasyCom dataset, and the second, D2, contains the simulated version of EasyCom (D1). 

In EasyCom (D1), one participant (ID=2) was wearing the array while others were only wearing a close-talking headset microphone used as an approximate reference signal. Although a variety of metadata is provided in EasyCom, the tested beamformers only used the array's \acp{ATF}, the target \acp{DOA} relative to the wearer (ID=2) and head-tracking data for each talker obtained from an optical motion capture system. The talkers' close-talking headset microphone was only used as the reference signal during calculation of intrusive metrics in D1. 

The D1 dataset embodies real-world realism and complexity. However, in such recordings, it is impossible to define the ground truth signals (direct-path target) at the array, that are normally required as reference signals for the calculation of instrumental (objective) intrusive metrics. Although close-talking headset microphone recordings for each talker are provided in  EasyCom, their use in calculation of intrusive evaluation metrics would introduce reference signal errors. Such errors may be caused by, for example, sensor noise, cross-talk leakage from other talkers, the frequency response of the close-talking microphone and its position relative to the mouth, as well as the potential inaccuracy in time and level alignments. Hence, our evaluation also uses the D2 simulated dataset where the true reference signals (noise-free direct-path target at the array) are available, at the cost of degraded realism compared to D1. The access to the ground truth noise signals for the data-driven dictionary is also important and only available in D2.  

In terms of the scenarios, both D1 and D2 share the same room dimensions of approximately $7\times6\times\SI{3}{\m}$, reverberation time RT$_{60}=\SI{645}{\ms}$ and ten fixed loudspeakers distributed across the room playing uncorrelated, restaurant-like ambient noise. The D2 dataset was simulated using the TASCAR software platform \citep{grimmToolboxRenderingVirtual2019a} with the presence of a table and the walls of approximate dimensions. The talkers' source signals in D2 were the D1 talkers' close-talking headset microphone signals, denoised using the CEDAR DNS Two plugin. Unlike D1 where the \acp{ATF} can slightly vary due to different persons wearing the array, the simulations in D2 were based on the same measured \acp{ATF} provided in \citep{Donley2021a}. The D2 dataset also used the same head-tracking data and voice activity labels as provided in EasyCom. 

In order to evaluate the effect of noise suppression on the localized interferer(s) such as the interfering talkers, the scenarios in EasyCom are divided into $\SI{6}{\s}$ segments such that the following conditions are met:

\begin{enumerate}
    \item Only one of the talkers is assumed as the target.
    \item The target onset occurs after $\SI{2}{\s}$ based on the voice activity metadata.
    \item The wearer participant (ID=2) is not active.
    \item The `waiter' participant (ID=1) whose \ac{DOA} and reference signal are unavailable is not the target.
\end{enumerate}
The segments that do not meet these constraints are discarded. Note that while the selected segments share these constraints, they can still differ in talker, utterance, number of active sources and \acp{DOA}. For the visualization of these results, the selected segments were categorized according to the number of active sources per segment, denoted as $\nTalker$ = $[1, 2, 3]$. The same segments from both D1 and D2 were used in the evaluation.
%%%%%%%%%%%%%%%%%%GOT THIS FAR

\begin{figure*}
    \centering
    \includegraphics[scale=.53]{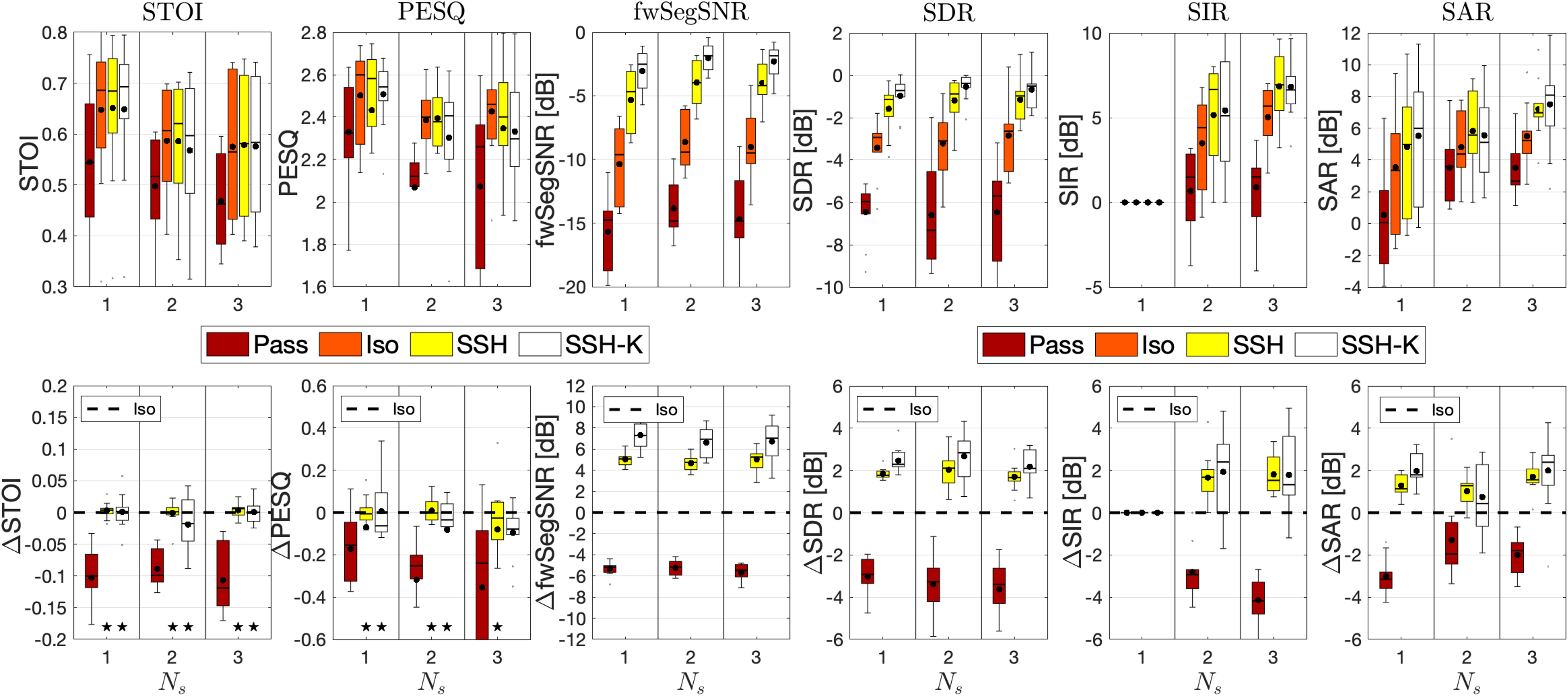}
    \caption{The absolute (top row) and relative (bottom row) STOI, PESQ, fwSegSNR, SDR, SIR and SAR scores of methods for $\nTalker=[1,2,3]$ in D1 dataset.}
    \label{fig:results_D1}
\end{figure*}

\subsection{Methods}

The evaluation compares the proposed SS-Hybrid method, including parametric and data-driven dictionaries, denoted as `$\SSH$' and `$\SSHK$', respectively, with the baseline superdirective beamformer denoted as `Iso', and the passthrough signal at the reference channel denoted as `Pass'. 

To avoid spatial aliasing due to the array geometry, data was lowpass filtered and down-sampled to $\SI{10}{\kHz}$ sample rate. The \ac{STFT} used a $\SI{16}{\ms}$ time-window and $\SI{8}{\ms}$ step. Microphone 2 (mid-front in Fig.~\ref{fig:EasyCom_array}) was selected as the reference channel. Empirical analysis showed that no condition number limiting on $\covMat$ in \eqref{eq:MVDRweights} was needed for any of the beamformers and therefore $\scalarLoading=0$ in \eqref{eq:robustCovMat}. The steering vector at each time-frame was updated with the \ac{RTF} associated with the closest direction to the target \ac{DOA} from the set of \ac{ATF} measurement points described in Section~\ref{ssec:array}. The settings associated with each method were as follows. 

\subsubsection{Iso (baseline superdirective)}
\label{sssec:methods_iso}

The directions set $\DOAset$ in \eqref{eq:Riso} used all the available \ac{ATF} measurement points described in Section~\ref{ssec:array} with quadrature weighting as in \eqref{eq:quadWeights}.   

\subsubsection{$\SSH$ (SS-Hybrid with parametric dictionary)}
\label{sssec:methods_ssh}

The parametric dictionary contained an isotropic model $\covMat\Iso$ as in Iso, an identity model $\covMat\Identity$, and $300$ unimodal anisotropic models $\covMat\Aniso$ as in \eqref{eq:dict_aniso} with all possible pairs between sixty different $\az\uPeak=\{0, 6, \dots, 354\}\deg$ and five different $\powDyn=\{8, 16, 24, 32, 40\}$~dB. This results in total of $\nModel=302$ models. The set of possible steering directions was limited to $\pm90\deg$ azimuth and $\pm30\deg$ elevation resulting in total of $\nDir=217$ possible steering directions. Considering $\nFreq=81$ frequency bands based on the sample rate and the \ac{STFT} settings as well as $\nMic=6$ channels based on the array configuration, the dictionary $\dict$ in \eqref{eq:dictionary} becomes a $302\times81\times6\times217$ table containing approximately $32$ million complex numbers. The time constant for the smoothness factor, $\smoothness$, in \eqref{eq:MethodsCovMatEst} was empirically set to $\timeConst=\SI{80}{\ms}$. 

\subsubsection{$\SSHK$ (SS-Hybrid with data-driven dictionary)}
\label{sssec:methods_sshk}

Apart from the content of the dictionary, the settings for $\SSHK$ were the same as for $\SSH$ unless stated otherwise. The set of ground truth \ac{NCM} estimates $\covRSet_{\iFreq} = \{\covMat\Oracle(\TF)\}$ in Section~\ref{sssec:clustering} were obtained using \eqref{eq:NCM_oracle} on half of the segments with $\nTalker=1$ in D2, which resulted in approximately $\nData=7500$ time-frames. The segments included in the training set were excluded from the evaluation results. The k-means clustering had random centroid initialization and used $\nCluster = \nModel = 302$ in order to have the same number of models in both data-driven and parametric dictionaries. The Iso model was additionally added to the data-driven dictionary as a requirement in Hybrid-\ac{MVDR}.

\subsection{Metrics}

The single-channel output of each method was compared with the reference signal for six intrusive metrics: \ac{STOI} \citep{Taal2011}, \ac{PESQ} \citep{ITU_T_P862,Rix2001}, \ac{fwSegSNR} \citep{Tribolet1978,Hu2008b}, \ac{SDR} \citep{vincentPerformanceMeasurementBlind2006a}, \ac{SIR} \citep{vincentPerformanceMeasurementBlind2006a} and \ac{SAR} \citep{vincentPerformanceMeasurementBlind2006a}. The VOICEBOX MATLAB toolbox \citep{Brookes1997} was used for \ac{STOI}, \ac{PESQ} and \ac{fwSegSNR} while the PEASS MATLAB toolbox \citep{Valentin2011,PEASS} was used for the \ac{SDR}, \ac{SIR} and \ac{SAR}. 

These metrics evaluate the performance of speech enhancement in terms of speech intelligibility (\ac{STOI}), speech quality (\ac{PESQ}) and the noise suppression (\ac{fwSegSNR}, \ac{SDR}, \ac{SIR}, \ac{SAR}). As for the reference signal used in the intrusive metric calculations, in the D1 dataset, the close-talking headset microphone signal was time- and level-aligned to the array `passthrough' signal at the reference channel using \texttt{sigalign} function from the VOICEBOX toolbox \citep{Brookes1997}. The true reference signal (noise-free direct-path target) was available in the D2 dataset. 

\subsection{Results: Real-recordings (D1)}
\label{ssection:results_D1}

Figure~\ref{fig:results_D1} shows the distributions of the absolute (top row) and the relative (bottom row) \ac{STOI}, \ac{PESQ}, \ac{fwSegSNR}, \ac{SDR}, \ac{SIR} and \ac{SAR} scores of the methods for varying $\nTalker=[1,2,3]$ using the real-recording D1 dataset. The boxes show the upper and lower inter-quartile range with mean and median indicated by black dot (•) and horizontal dash (-) markers, respectively. The whiskers extend to $1.5$ times of the inter-quartile range. In the bottom row plots, the presence of a star indicates no statistically significant difference from the baseline Iso beamformer (dashed line) according to paired t-test at $p=5\%$ level. The discussion of the results based on the category of metrics is as follows.

\subsubsection{D1 Speech Intelligibility (\ac{STOI})}
\label{sssec:D1_intelligibility}
As shown in Fig.~\ref{fig:results_D1} first column, all beamformers improved the \ac{STOI} compared to passthrough by a mean $\Delta\ac{STOI}=\SI{0.1}{}$. Both SS-Hybrid implementations ($\SSH$ and $\SSHK$) preserved the same $\ac{STOI}$ improvement as the baseline Iso.

\subsubsection{D1 Speech Quality (\ac{PESQ})}
\label{sssec:D1_quality}
The second column of Fig.~\ref{fig:results_D1} also shows the same improvement in $\ac{PESQ}$ scores with mean $\Delta\ac{PESQ}=[0.2, 0.3, 0.4]$ for all beamformers compared to passthrough in $\nTalker=[1,2,3]$ respectively, with an exception for $\SSHK$ in $\nTalker=3$ with slightly less $\Delta\ac{PESQ}$ than Iso and $\SSH$. 

\subsubsection{D1 Noise Suppression (\ac{fwSegSNR}, \ac{SDR}, \ac{SIR}, \ac{SAR})}
\label{sssec:D1_suppression}
The third to the sixth columns in Fig~\ref{fig:results_D1} show the suppression results for different definitions of noise. As expected, all beamformers significantly reduced the noise compared to the passthrough. Both versions of the proposed method significantly outperformed the baseline Iso by an average of $\SI{5}{\dB}$ for \ac{fwSegSNR} and $\SI{3}{\dB}$ for \ac{SDR}, \ac{SIR} and \ac{SAR}, in addition to the noise suppression achieved by Iso. The data-driven dictionary ($\SSHK$) suppressed noise $\SI{2}{\dB}$ more than the parametric dictionary ($\SSH$) due to the use of the data-driven \acp{NCM}.

\begin{figure}[t]
    \centering
    \includegraphics[scale=.43]{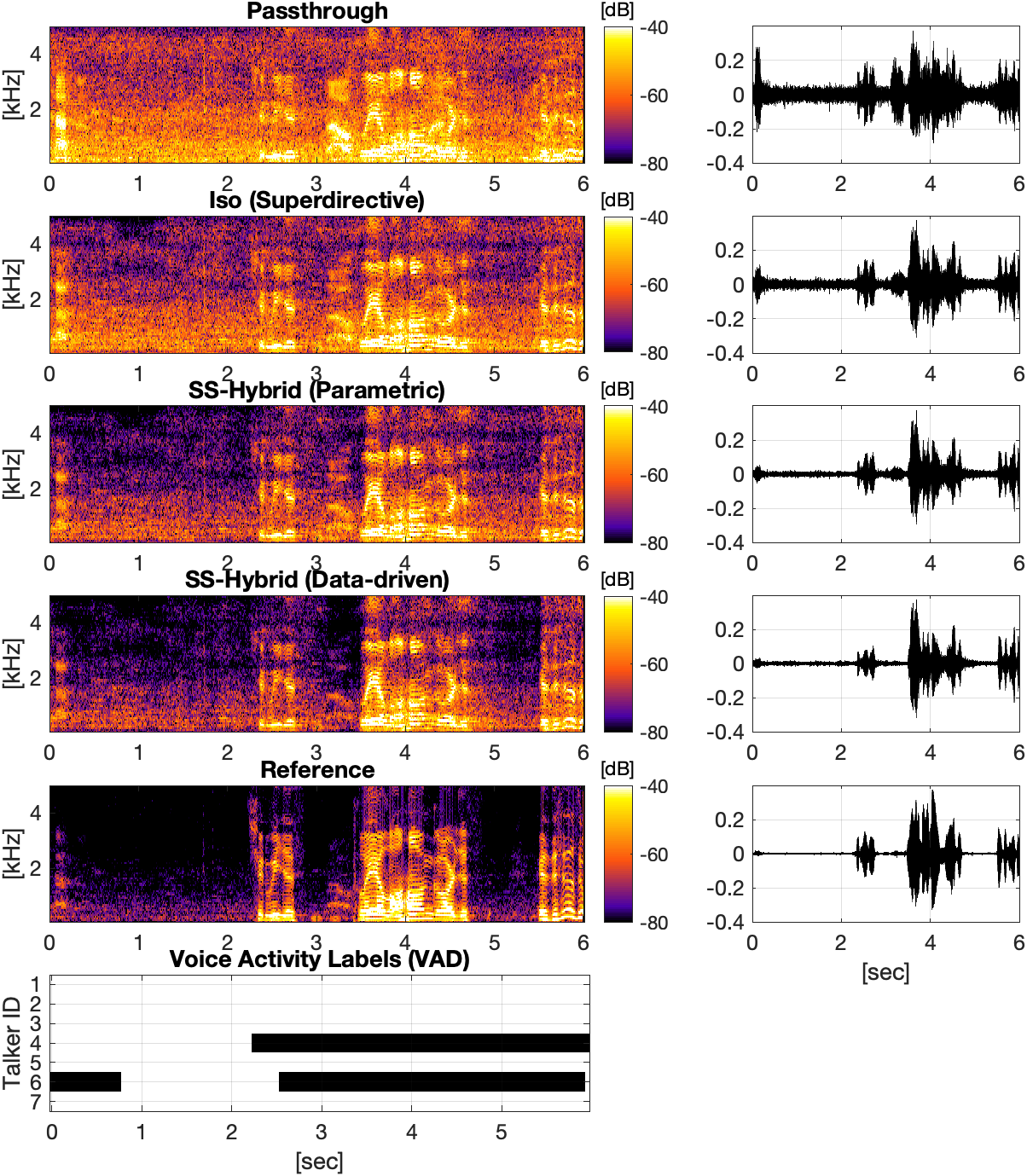}
    \caption{The output spectrogram and waveform for each method from an example trial with $\nTalker=2$ in real-recording D1 dataset.}
    \label{fig:stft}
\end{figure}

Figure~\ref{fig:stft} illustrates the spectrogram and waveform outputs of Pass, Iso, $\SSH$, $\SSHK$ and the Reference signals from an example in D1 with an  overlapping interferer talker ($\nTalker=2$). In line with the metrics discussed previously, it can be seen that the proposed SS-Hybrid method provides stronger noise suppression, while preserving the target signal as well as the baseline Iso. This example also shows the noise suppression superiority of data-driven ($\SSHK$) over parametric ($\SSH$) dictionaries for the proposed method. Audio demo examples for this and other cases from both D1 and D2 datasets are available in \citep{SSHybridDemo}.

\subsubsection{Musical Noise Removal}
\label{ssec:D1_suppression}

\begin{figure}[t]
    \centering
    \includegraphics[scale=.43]{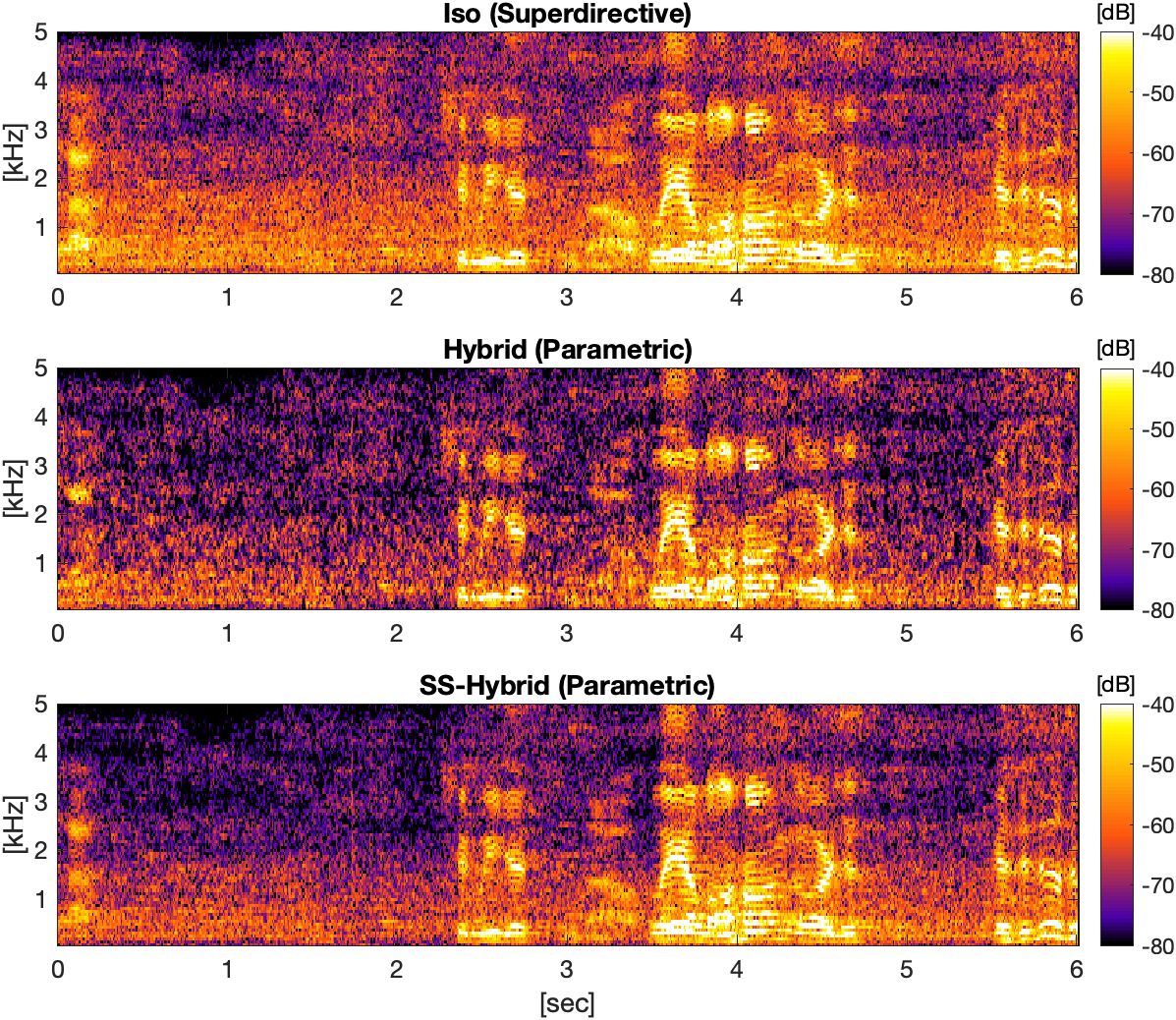}
    \caption{The spectrogram outputs of Iso, Hybrid and SS-Hybrid for the same example as in Fig.~\ref{fig:stft}.}
    \label{fig:musical_noise}
\end{figure}

As shown in the spectrograms of Iso, Hybrid and SS-Hybrid in Fig.~\ref{fig:musical_noise}, the musical noise in the Hybrid (middle row) is successfully removed by the inter-method wideband \ac{PCA} in SS-Hybrid (bottom row) while preserving its extra noise suppression compared to Iso (top row). 

\subsection{Results: Simulation (D2)}
\label{ssection:results_D2}

\begin{figure*}[t]
    \centering
    \includegraphics[scale=.53]{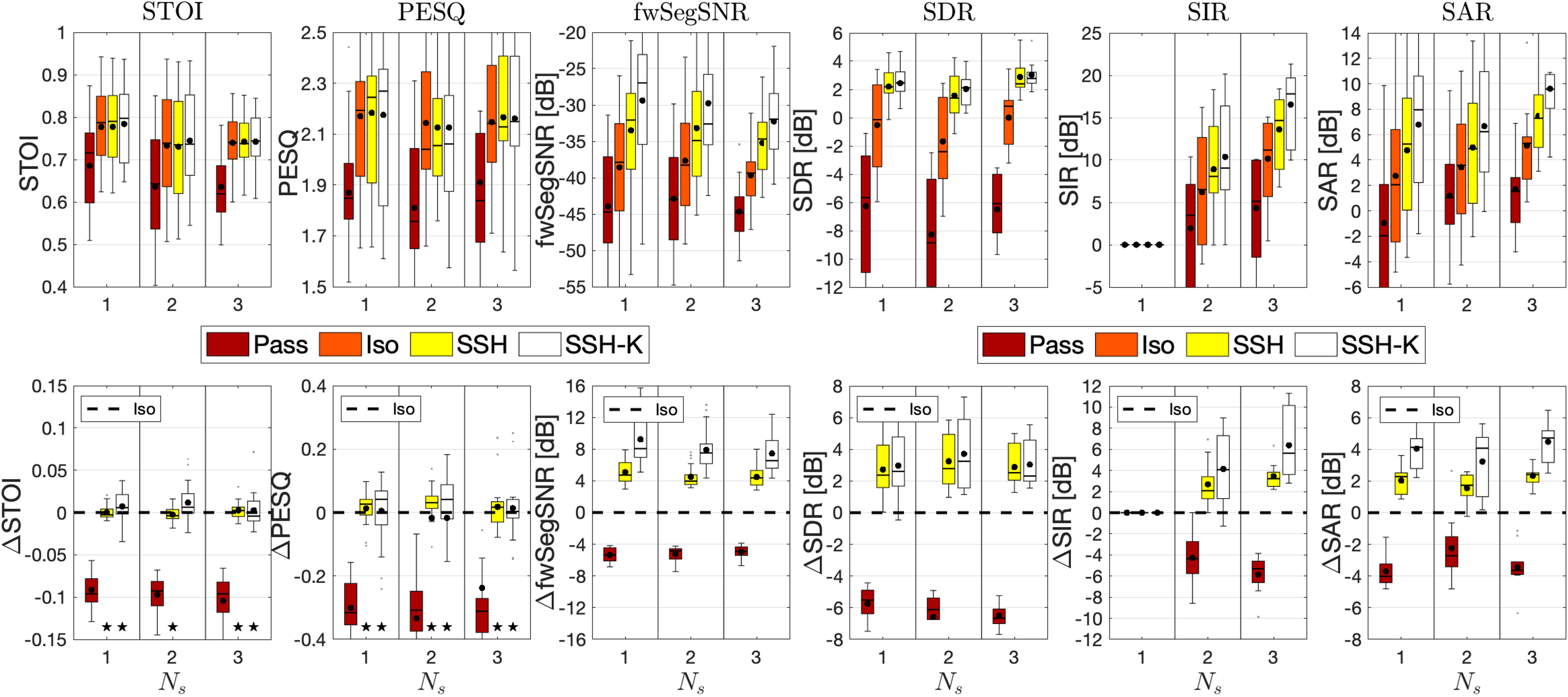}
    \caption{The absolute (top row) and relative (bottom row) \ac{STOI}, \ac{PESQ}, \ac{fwSegSNR}, \ac{SDR}, \ac{SIR} and \ac{SAR} scores of methods for $\nTalker=[1,2,3]$ in D2 dataset.}
    \label{fig:results_D2}
\end{figure*}

Figure~\ref{fig:results_D2} shows the distributions of the absolute (top row) and the relative (bottom row) metrics for the methods for varying $\nTalker=[1,2,3]$ using the simulated D2 dataset. All beamformers improved all metrics compared to passthrough as expected.

\subsubsection{D2 Speech Intelligibility (\ac{STOI})}
\label{sssec:D2_intelligibility}

Both variations of the proposed SS-Hybrid method ($\SSH$ and $\SSHK$) at least maintain the same intelligibility improvement as in the baseline Iso with the exception of data-driven $\SSHK$ providing marginally higher $\Delta\ac{STOI}$ improvement for $\nTalker=2$. 

\subsubsection{D2 Speech Quality (\ac{PESQ})}
\label{sssec:D2_quality}

Similar to the \ac{STOI} results, both $\SSH$ and $\SSHK$ preserve the same \ac{PESQ} improvement achieved by the baseline beamformer with an average $\Delta\ac{PESQ}=0.3$ for all beamformers compared to the passthrough.

\subsubsection{D2 Noise Suppression (\ac{fwSegSNR}, \ac{SDR}, \ac{SIR}, \ac{SAR})}
\label{sssec:D2_suppression}

As expected and similar to the D1 dataset, the two versions of the proposed method ($\SSH$ and $\SSHK$) significantly outperform the baseline Iso in all noise suppression metrics by an average of $\Delta\ac{fwSegSNR}=\SI{6}{\dB}$ and $\SI{3}{\dB}$ for \ac{SDR}, \ac{SIR} and \ac{SAR}. The SS-Hybrid with data-driven dictionary ($\SSHK$), compared to the parametric dictionary ($\SSH$), provides mean improvements in noise suppression of $\SI{4}{\dB}$ in \ac{fwSegSNR} and $\SI{2}{\dB}$ in \ac{SDR}, \ac{SAR} and \ac{SIR} for the reasons discussed in Section~\ref{sssec:D1_suppression}. The good performance of the proposed methods shows robustness to an increase in the number of active sources ($\nTalker$) and is consistent between the real-recording (D1) and simulated (D2) datasets.

\section{Conclusions}
\label{sec:conclusion}

A method of beamforming is proposed for multi-channel speech enhancement that is adaptive and signal-dependent while being comparable in terms of computational cost to static and signal-independent beamformers. The proposed beamformer uses a dictionary of pre-calculated weights. The method operates in two stages where, in the first stage, multiple \ac{MVDR} beamformers are implemented using a weight dictionary with entries corresponding to different noise covariance models. In the second stage, a wideband inter-method \ac{PCA} is applied to the outputs of both static isotropic and adaptive Hybrid-model beamformers per time-frame to remove any `musical noise' caused by the Hybrid $\ac{MVDR}$ beamforming. Two alternative dictionary designs were investigated: a parametric  dictionary using isotropic and horizontally unimodal anisotropoic noise field isotropy assumptions, and a data-driven dictionary using k-means clustering and ground truth-aware estimates of the noise covariance matrices from a simulated dataset. In the context of augmented hearing and augmented reality audio, evaluations were performed using real recordings of `cocktail party' scenarios with approximate reference signals, and simulations with accurate reference signals for a $6$-channel head-worn microphone array. The results for both real recordings and simulations showed significant performance benefits of the proposed method compared to the baseline superdirective beamformer. In terms of noise suppression, average improvements of $\SI{6}{\dB}$ \ac{fwSegSNR} and $\SI{3}{\dB}$ \ac{SDR}, \ac{SIR} and \ac{SAR} were found, while improvements in speech intelligibility (\ac{STOI}) and quality (\ac{PESQ}) were at least as good as the baseline superdirective beamformer. The use of a data-driven dictionary design, compared to the parametric dictionary, in the proposed method provided $1$ to $\SI{2}{\dB}$ more noise suppression with a corresponding improvement in the speech intelligibility and quality.
\appendix
The plane-wave representation of the array signal in \eqref{eq:PWD}  with $\nPW$ \acp{PW} can be re-written by separating the target \ac{PW} and the remaining $\nPW-1$ \acp{PW} as 

\begin{equation}
\stftArray_{\iMic}(\TF) = \stftTarget_{\iMic}(\TF)+\sum_{\iPW=1}^{\nPW-1}\stftPW_{\iMic,\iPW}(\TF) 
+ \stftWhite_{\iMic}(\TF),
\label{eq:PWDtarget}
\end{equation}
where $\stftTarget_{\iMic}=\stftPW_{\iMic,\iPW}\in\Complex$ is the $\iPW$-th \ac{PW} associated with the target direction $\direction\Target$. Since formulations for the remaining of this section will be all for a reference microphone, the microphone index $\iMic$ will be omitted for notational simplicity. Equation \eqref{eq:PWDtarget} can be written for the full-band spectrum at a reference microphone as

\begin{align}
\check{\stftArrayVec}(\iFrame) & = \check{\stftTargetVec}(\iFrame)+\check{\stftPWVec}(\iFrame) 
+ \check{\stftWhiteVec}(\iFrame)\\
& = \check{\stftTargetVec}(\iFrame)+\check{\stftNoiseVec}(\iFrame)
\label{eq:PWDvec}
\end{align}
where $\check{\stftArrayVec}\in\Complex^{1\times\nFreq}$, $\check{\stftTargetVec}$, $\check{\stftPWVec}=\sum_{\iPW=1}^{\nPW-1}\check{\stftPWVec}_{\iPW}$ and $\check{\stftWhiteVec}$ are the full-band spectrum of the array, target \ac{PW} and the remaining \acp{PW} while  $\check{\stftNoiseVec}=\check{\stftPWVec}+\check{\stftWhiteVec}$ denotes the residual noise spectrum including the mixture of non-Target \acp{PW} and the sensor noise.

The \ac{MVDR} spectral output of Iso and Hybrid with steering direction $\direction\Target$ can be written as 

\begin{align}
\check{\stftOutVec}\Iso(\iFrame)&=\check{\stftTargetVec}(\iFrame)+\check{\stftNoiseVec}\Iso(\iFrame),\label{eq:IsoPW}\\
\check{\stftOutVec}\Hyb(\iFrame)&=\check{\stftTargetVec}(\iFrame)+\check{\stftNoiseVec}\Hyb(\iFrame),\label{eq:HybPW}
\end{align}
where $\check{\stftNoiseVec}\Iso$ and  $\check{\stftNoiseVec}\Hyb$ are the residual noise in the output of Iso and Hybrid-\ac{MVDR} beamformers assuming the target \ac{PW} $\check{\stftTargetVec}$ is preserved due to the distortionless response constraint in the \ac{MVDR} formulation. Note that $\check{\stftNoiseVec}\Iso$ and  $\check{\stftNoiseVec}\Hyb$ are the frequency-dependent scaled versions of $\check{\stftNoiseVec}$ in \eqref{eq:PWDvec} where the scaling coefficients are determined by the narrowband complex-valued beam patterns. The $\check{\stftTargetVec}$, $\check{\stftNoiseVec}\Iso$ and $\check{\stftNoiseVec}\Hyb$ are expected and assumed to be uncorrelated. 

Substituting \eqref{eq:IsoPW} and \eqref{eq:HybPW} in \eqref{eq:MethodsArray} gives
\begin{equation}
    \stftOutMat = \begin{bmatrix}
    \check{\stftTargetVec}\\
    \check{\stftTargetVec}
    \end{bmatrix}
    +
    \begin{bmatrix}
    \check{\stftNoiseVec}\Hyb\\
    \check{\stftNoiseVec}\Iso
    \end{bmatrix},\label{eq:MethodsArraySplit}
\end{equation}
where time-frame dependency $(\iFrame)$ is omitted for notational simplicity. 
The inter-method wideband covariance matrix $\covMat\Out\in\Complex^{2\times2}$ in \eqref{eq:MethodsCovMat} is then re-written as

\begin{equation}
\begin{split}
    \covMat\Out
    &=\expectOp\{\stftOutMat\stftOutMat\hermitianOp\}
    \\
    &=\expectOp\{
    \begin{bmatrix}
        \check{\stftTargetVec}\check{\stftTargetVec}\hermitianOp & \check{\stftTargetVec}\check{\stftTargetVec}\hermitianOp\\
       \check{\stftTargetVec}\check{\stftTargetVec}\hermitianOp & \check{\stftTargetVec}\check{\stftTargetVec}\hermitianOp
    \end{bmatrix}
    \}
    +
    \expectOp\{
    \begin{bmatrix}
        \check{\stftNoiseVec}\Hyb\check{\stftNoiseVec}\Hyb\hermitianOp & \check{\stftNoiseVec}\Hyb\check{\stftNoiseVec}\Iso\hermitianOp\\
       \check{\stftNoiseVec}\Iso\check{\stftNoiseVec}\Hyb\hermitianOp & \check{\stftNoiseVec}\Iso\check{\stftNoiseVec}\Iso\hermitianOp
    \end{bmatrix}
    \}
    \\
    &
    + 
    \expectOp\{
    \begin{bmatrix}
        \check{\stftTargetVec}\check{\stftNoiseVec}\Hyb\hermitianOp & \check{\stftTargetVec}\check{\stftNoiseVec}\Iso\hermitianOp\\
       \check{\stftTargetVec}\check{\stftNoiseVec}\Hyb\hermitianOp & \check{\stftTargetVec}\check{\stftNoiseVec}\Iso\hermitianOp
    \end{bmatrix}
    \}
    + 
    \expectOp\{
    \begin{bmatrix}
        \check{\stftNoiseVec}\Hyb\check{\stftTargetVec}\hermitianOp & \check{\stftNoiseVec}\Hyb\check{\stftTargetVec}\hermitianOp\\
       \check{\stftNoiseVec}\Iso\check{\stftTargetVec}\hermitianOp & \check{\stftNoiseVec}\Iso\check{\stftTargetVec}\hermitianOp
    \end{bmatrix}
    \}.
\label{eq:MethodsCovMatSplit}
\end{split}
\end{equation}
Since $\check{\stftTargetVec}$, $\check{\stftNoiseVec}\Iso$ and $\check{\stftNoiseVec}\Hyb$ are uncorrelated, \eqref{eq:MethodsCovMatSplit} can be simplified to

\begin{align}
\begin{split}
    \covMat\Out
    &=\lVert\check{\stftTargetVec}\rVert^{2}
    \AllOnesMat
    +
    \begin{bmatrix}
        \lVert\check{\stftNoiseVec}\Hyb\rVert^{2} & 0\\
       0 & \lVert\check{\stftNoiseVec}\Iso\rVert^{2}
    \end{bmatrix}
    \\
    &
    =\lVert\check{\stftTargetVec}\rVert^{2}
    \AllOnesVec\AllOnesVec\hermitianOp
    +
    \sigmaPlus
    \eye
    +
    \sigmaMinus
    \signatureMat,
\label{eq:MethodsCovMatSplit2}
\end{split}\\
\mathrm{where}\quad \sigmaPlus &= (\frac{\lVert\check{\stftNoiseVec}\Hyb\rVert^{2}+\lVert\check{\stftNoiseVec}\Iso\rVert^{2}}{2}),
\\
\mathrm{and}\quad \sigmaMinus &= (\frac{\lVert\check{\stftNoiseVec}\Hyb\rVert^{2}-\lVert\check{\stftNoiseVec}\Iso\rVert^{2}}{2}),
\label{eq:n2}
\end{align}
where $\AllOnesMat=\AllOnesVec\AllOnesVec\hermitianOp\in\Real^{2\times2}$ is the all-one matrix with $\AllOnesVec=[1,1]\transposeOp$ being all-one column-vector, $\eye=\diag([1,1])\in\Real^{2\times2}$ is the identity matrix and $\signatureMat=\diag([1,-1])\in\Real^{2\times2}$ is the signature matrix. Hence, the signal subspace in the inter-method \ac{PCA} is expected to dominantly include $\check{\stftTargetVec}$ that is the most correlated component between the $\check{\stftOutVec}\Hyb$ and $\check{\stftOutVec}\Iso$ and avoids the least correlated component between the two that is the musical noise only present in $\check{\stftNoiseVec}\Hyb$.

% \input{wip}

%\input{appendices}
%--\input{acknowledgment}
%\todo[inline]{references-bibfile.tex starts here}
% \input{references-bibfile}

\bibliographystyle{IEEEtran}
\bibliography{bibtex/bib/sapstrings.bib,bibtex/bib/sapref.bib,bibtex/bib/extras.bib}
% \begin{thebibliography}{99}
% \input{bibitems}
% \input{bibtex/bib/saprefs}
%\end{thebibliography}

%\input{references}
%--
%\vskip -2\baselineskip plus -1fil
%\vskip 0pt plus -1fil

\begin{IEEEbiography}[{\includegraphics[width=1in,height=1.25in,clip,keepaspectratio]{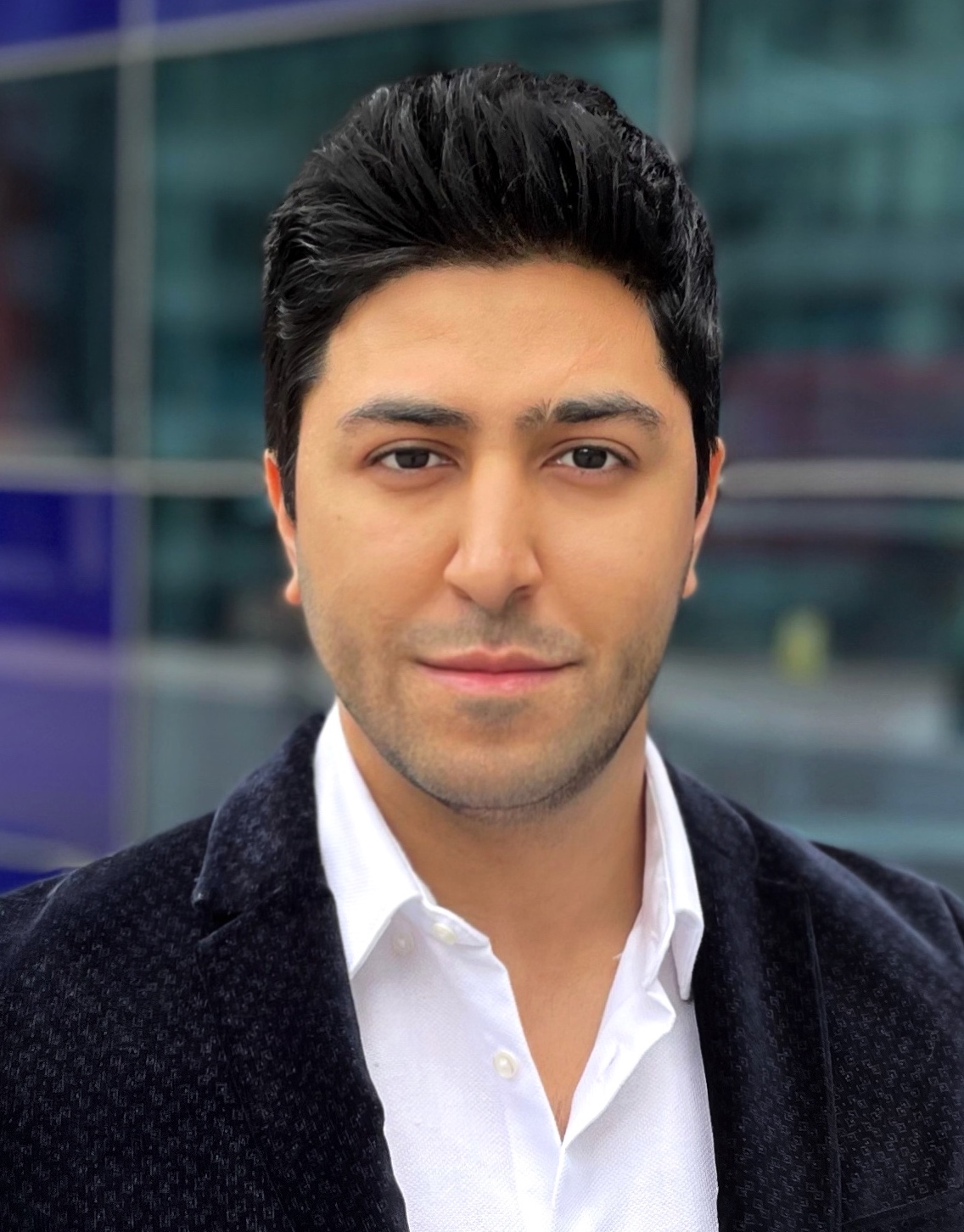}}]{Sina Hafezi}
is a post-doctoral Research Associate at Imperial College London. He received the BEng degree in Electronic Engineering in 2012 and the MSc degree in Digital Signal Processing in 2013 both from Queen Mary, University of London, UK. He worked in the Centre for Digital Music as a researcher and software engineer on autonomous multitrack mixing systems, which led to patent and spin-out company. In 2018, he received his PhD in acoustic source localisation using spherical microphone arrays at Imperial College London, UK. He spent 2.5 years at Silixa as Senior Signal Processing Engineer developing algorithms and software for Distributed Acoustic Sensing systems. In 2021, he re-joined Imperial College London, where he has contributed to academic and industrial projects on hearing aids and spatial audio. His research interests are microphone array processing, spatial audio rendering, beamforming, source localisation and room acoustic modelling with applications for augmented and virtual reality.
\end{IEEEbiography}

%\vskip -2\baselineskip plus -1fil
\vskip 0pt plus -1fil

\begin{IEEEbiography}[{\includegraphics[width=1in,height=1.25in,clip,keepaspectratio]{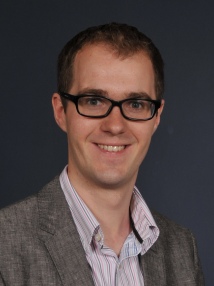}}]{Alastair H. Moore}
(M'13) is  a  Research Fellow at  Imperial  College  London  and  spatial  audio  consultant  with  Square  Set  Sound.  He  received  the M.Eng. degree in Electronic Engineering with Music Technology Systems in 2005 and the Ph.D. degree in 2010, both from the University of York, York, U.K. He  spent  3  years  as  a  Hardware  Design  Engineer for Imagination Technologies PLC designing digital radios  and  networked  audio  consumer  electronics products. In 2012, he joined Imperial College, where he has contributed to a series of projects in the field of speech and audio processing applied to voice over IP, robot audition, and hearing  aids.  Particular  topics  of  interest  include  microphone  array  signal processing, modeling and characterization of room acoustics, dereverberation, and  spatial  audio  perception.  His  current  research  is  focused  on  signal processing for moving, head-worn microphone arrays.
\end{IEEEbiography}

%\vskip -2\baselineskip plus -1fil
\vskip 0pt plus -1fil

\begin{IEEEbiography}[{\includegraphics[width=1in,height=1.25in,clip,keepaspectratio]{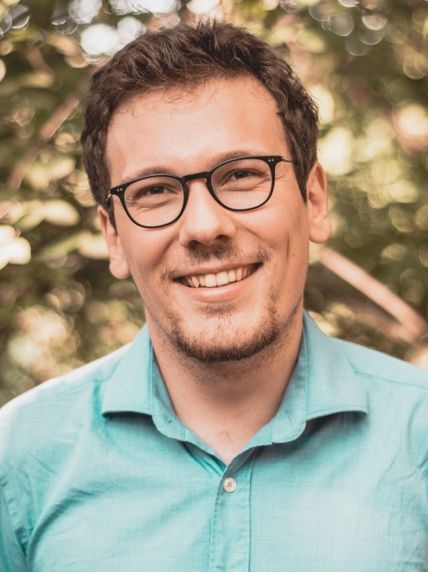}}]{Pierre H. Guiraud} 
is a post-doctoral Research Associate at Imperial College
London. After enrolling in a double degree program, he received a MSc
in Engineering from the Ecole Centrale de Lille, France, as well as a
Master in Engineer Acoustic from the Technical University of Denmark in
Copenhagen in 2017. His master thesis focused on ambisonic sound
reproduction in collaboration with the company Kahle Acoustics. He went
on to pursue a PhD at the IEMN in Lille, France, about thethermoacoustic sound generation in porous metamaterials.  This work was 
sponsored by Thales Underwater Systems and done in partnership with
CINTRA Singapore. He obtained his PhD in 2020 and joined Imperial
College London in 2021 working on several projects. Within the Speech and Audio Processing group, he worked on the SPEAR Challenge which is an IEEE challenge about
speech enhancement for Virtual/Augmented Reality in partnership with Meta
Reality Labs Research. Within the ELO-SPHERES
project, he worked on how to
improve binaural intelligibility metrics in real
environments using deep learning, in collaboration with University College London. Lastly, in the Audio Experience Design research group, he was in charge of an experiment on spatial audioperception in a virtual reality environment.
\end{IEEEbiography}

%\vskip -2\baselineskip plus -1fil
\vskip 0pt plus -1fil

\begin{IEEEbiography}[{\includegraphics[width=1in,height=1.25in,clip,keepaspectratio]{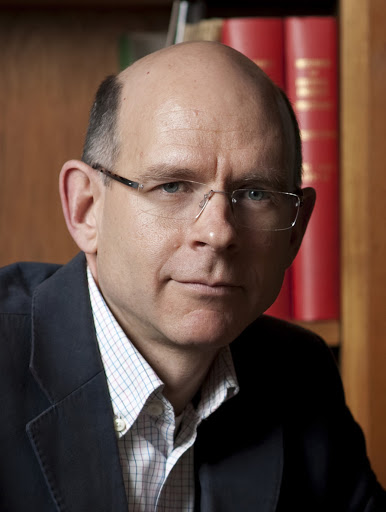}}]{Patrick A. Naylor}
(M’89, SM’07, F’20) is Professor of Speech and Acoustic Signal Processing at Imperial  College  London.  He  received  the  BEng  degree in  Electronic  and  Electrical  Engineering  from  the University  of  Sheffield,  UK,  and  the  PhD  degree from  Imperial  College  London,  UK.  His  research interests are in speech, audio and acoustic signal processing. His current research addresses microphone array   signal   processing,   speaker   diarization,   and multichannel speech enhancement for application to binaural  hearing  aids  and  robot  audition.  He  has also  worked  on  speech  dereverberation  including  blind  multichannel  system identification  and  equalization,  acoustic  echo  control,  non-intrusive  speech quality  estimation,  and  speech  production  modelling  with  a  focus  on  the analysis of the voice source signal. In addition to his academic research, he enjoys several collaborative links with industry. He is currently a member of the Board of Governors of the IEEE Signal Processing Society and President of  the  European  Association  for  Signal  Processing  (EURASIP).  He  was formerly Chair of the IEEE Signal Processing Society Technical Committee on Audio and Acoustic Signal Processing. He has served as an associate editor of  IEEE  Signal  Processing  Letters  and  is  currently  a  Senior  Area  Editor  of IEEE Transactions on Audio Speech and Language Processing.
\end{IEEEbiography}

%\vskip -2\baselineskip plus -1fil
\vskip 0pt plus -1fil

\begin{IEEEbiography}[{\includegraphics[width=1in,height=1.25in,clip,keepaspectratio]{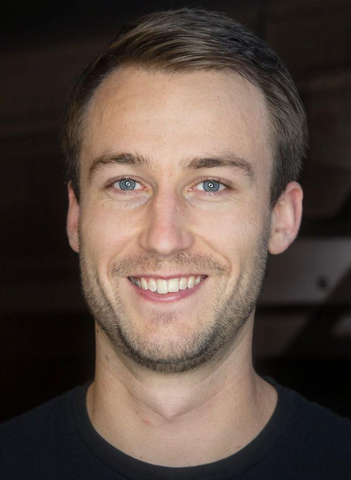}}]{Jacob Donley}
is a research scientist at Meta Reality Labs Research in Redmond, WA with previous experience lecturing in Signals and Systems at the University of Wollongong (UOW) and Engineering at Western Sydney University. He holds a Bachelor of Engineering Honours (Computer) (UOW), a Ph.D. (UOW) with research focused on Digital Signal Processing (DSP) aimed at improving the reproduction of personal sound in shared environments, and a graduate certificate in Artificial Intelligence (AI) from Stanford University. Jacob’s research interests are in signal processing, speech enhancement, machine learning, array processing (microphone and loudspeaker), beamforming, and multi-zone sound field reproduction. Jacob has received awards and scholarships from Telstra Corporation Limited, the Australian Department of Education and Training, and the University of Wollongong. He is also an active member of the IEEE and IEEE Signal Processing Society.

\end{IEEEbiography}

%\vskip -2\baselineskip plus -1fil
\vskip 0pt plus -1fil

\begin{IEEEbiography}[{\includegraphics[width=1in,height=1.25in,clip,keepaspectratio]{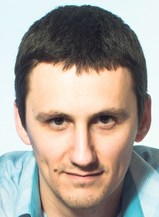}}]{Vladimir Tourbabin}
(M’16) is a Research Science Manager at Meta. He received the B.Sc. degree (summa cum laude) in materials science
and engineering in 2005, the M.Sc. degree (cum laude) in electrical and computer engineering in 2011, and the PhD degree in electrical and computer engineering in 2016. All three degrees are from Ben-Gurion University of the Negev, Israel. After the graduation, he has joined the General Motors’ Advanced Technical Center in Israel, to work on microphone array processing solutions for speech recognition. Since 2017, Dr. Tourbabin is with Reality Labs Research @ Meta (formerly known as Facebook Reality Labs) working on research and advanced development of audio signal processing technologies for augmented and virtual reality applications.
\end{IEEEbiography}

%\vskip -2\baselineskip plus -1fil
\vskip 0pt plus -1fil

\begin{IEEEbiography}[{\includegraphics[width=1in,height=1.25in,clip,keepaspectratio]{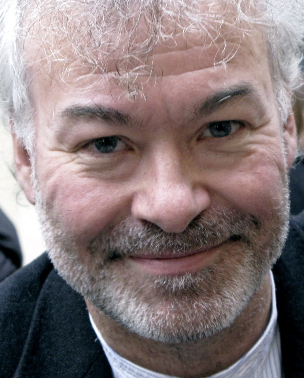}}]{Thomas Lunner's}
career has focused on man-machine issues related to the correlation between hearing and cognition. He was the first scientist to convincingly show the importance of cognitive ability in relation to being able to recognize speech in adverse listening conditions. He also demonstrated how working memory was a significant factor for the selection of optimum signal processing in hearing aids to be fitted to a particular individual. These studies led to increasing cooperation with other research groups globally, and the establishment of cognitive hearing science as a research field of its own. Dr. Lunner’s Ph.D. research in the mid-1990s resulted in patented signal processing algorithms which led to the development of the first digital hearing aid manufactured by Oticon. The core of this project was a digital filter bank which provided the necessary tuning flexibility with an equally important low power consumption. The filter bank was used in several successive hearing aid models in the years that followed, fitted to millions of hearing aid users worldwide. Two of the models were awarded the European Union’s prestigious technology prize, the IST Grand Prize in 1996 and in 2003. Currently, he leads research in Superhuman Hearing at Meta Reality Labs Research.
\end{IEEEbiography}

%\vskip -2\baselineskip plus -1fil
%\vskip 0pt plus -1fil

%\showdimensions %macro only defined for non-final mode

% multicols not being used
% \end{multicols} %just before \end{document}
\end{document}